\documentclass[transmag]{IEEEtran}

\def\BibTeX{{\rm B\kern-.05em{\sc i\kern-.025em b}\kern-.08em T\kern-.1667em\lower.7ex\hbox{E}\kern-.125emX}}

\ifCLASSINFOpdf
\usepackage[pdftex]{graphicx}

\else

\fi

\usepackage[cmex10]{amsmath}
\usepackage{amssymb}   
\usepackage{amsxtra}
\usepackage{amscd}
\usepackage{amsthm}
\usepackage{textcomp}
\usepackage{graphicx}

\usepackage{balance}

\usepackage{array}  

\usepackage{multirow}  

\usepackage{amsmath}
\usepackage{cite}

\usepackage{url}

\usepackage{color,soul} 
\soulregister\cite7
\soulregister\ref7
\soulregister\pageref7

\begin{document}
	
	\title{{\huge Reconfigurable Intelligent Surfaces for Doppler Effect and Multipath Fading Mitigation}}

	
	%
	%
	%
	
	\author{Ertugrul~Basar,~\IEEEmembership{Senior Member,~IEEE} 
	
		\thanks{E. Basar is with the Communications Research and Innovation Laboratory (CoreLab), Department of Electrical and Electronics Engineering, Ko\c{c} University, Sariyer 34450, Istanbul, Turkey. e-mail: ebasar@ku.edu.tr}
		\thanks{Codes available at https://corelab.ku.edu.tr/tools }
	}

\IEEEtitleabstractindextext{\begin{abstract}
Extensive research has already started on 6G and beyond wireless technologies due to the envisioned new use-cases and potential new requirements for future wireless networks. Although a plethora of modern physical layer solutions have been introduced in the last few decades, it is undeniable that a level of saturation has been reached in terms of the available spectrum, adapted modulation/coding solutions and accordingly the maximum capacity. Within this context, communications through reconfigurable intelligent surfaces (RISs), which enable novel and effective functionalities including wave absorption, tuneable anomalous reflection, and reflection phase modification, appear as a potential candidate to overcome the inherent drawbacks of legacy wireless systems. The core idea of RISs is the transformation of the uncontrollable and random wireless propagation environment into a reconfigurable communication system entity that plays an active role in forwarding information. In this paper, the well-known multipath fading phenomenon is revisited in mobile wireless communication systems, and novel and  unique solutions are introduced from the perspective of RISs. The feasibility of eliminating or mitigating the multipath fading effect stemming from the movement of mobile receivers is also investigated by utilizing the RISs. It is shown that rapid fluctuations in the received signal strength due to the Doppler effect can be effectively reduced by using the real-time tuneable RISs. It is also proven that the multipath fading effect can be totally eliminated when all reflectors in a propagation environment are coated with RISs, while even a few RISs can significantly reduce the Doppler spread as well as the deep fades in the received signal for general propagation environments with several interacting objects.

		\end{abstract}
		

}

\maketitle

\section{Introduction}

\IEEEPARstart{F}{ifth} generation (5G) systems have major  three  use-cases with diverse requirements, namely enhanced mobile broadband (eMBB),  ultra-reliable and low-latency communications (URRLC), and massive machine type communications (mMTC). Although the 5G standard exploits promising physical layer (PHY) technologies including massive multiple-input multiple-output (MIMO) systems, millimeter wave (mmWave) communications, and multiple orthogonal frequency division multiplexing (OFDM) numerologies, it is does not contain revolutionary ideas in terms of PHY layer solutions. From this perspective, researchers have already started research on beyond 5G, or even 6G technologies of 2030 and beyond by exploring completely new PHY concepts. Even though future 6G technologies seem to be extensions of their 5G counterparts at this time \cite{6G},  potential new user requirements, use-cases, and networking trends of 6G \cite{Saad_2019} will bring more challenging problems in mobile wireless communication, which necessitate radically new communication paradigms, particularly at the PHY of next-generation radios. The envisioned new communication solutions must provide extremely high spectral and energy efficiencies along with ultra-reliability and ultra-security, and must have highly flexible structures to satisfy the challenging requirements of diverse users and applications. Although the intensive research efforts of the past two decades, these are still missing features in state-of-the-art systems and standards, and slowing down the progress of long-awaited wireless revolution.
 
Since the invention of modern wireless communications, network operators have been constantly struggling to build truly pervasive wireless networks that can provide uninterrupted connectivity and high quality-of-service (QoS) to multiple users and devices in the presence of harsh propagation environments \cite{Molisch}. The main reason of this phenomenon is the uncontrollable and random behavior of wireless propagation, which causes \\
\hspace*{0.5cm}i) \textit{deep fading} due to uncontrollable interactions of transmitted waves with surrounding objects and their destructive interference at the receiver,\\
\hspace*{0.5cm}ii) \textit{severe attenuation} due to path loss, shadowing, and non-line-of-sight (LOS) transmissions,\\
\hspace*{0.5cm}iii) \textit{inter-symbol interference} due to different runtimes of multipath components, and\\
\hspace*{0.5cm}iv) \textit{Doppler effect} due to the high mobility of users and/or surrounding objects. 

Although a plethora of modern PHY solutions, including adaptive modulation and coding, multi-carrier modulation, non-orthogonal multiple access, relaying, beamforming, and reconfigurable antennas, have been considered to overcome these challenges in the next several decades, the overall progress has been still relatively slow. The major reason of this relatively slow progress is explained by the following so-called undeniable fact: \textit{until the start of modern wireless communications (for decades), the propagation environment has been perceived as a randomly behaving entity that degrades the overall received signal quality and the communication QoS due to uncontrollable interactions of the transmitted radio waves with the surrounding objects}. In other words, communication pundits have been mainly focusing on transmitter and receiver ends of traditional wireless communication systems “for ages” while assuming that the wireless communication environment itself remains an uncontrollable factor and has usually a negative effect on the communication efficiency and the QoS. One of the main objectives of this paper is to challenge this view by exploiting the new paradigm of intelligent communication environments.

In recent years, reconfigurable intelligent surfaces (RISs) have been brought to the attention of the wireless research community to enable the control of wireless environments \cite{Di_Renzo_2019,Basar_Access_2019}. RISs are man-made surfaces of electromagnetic (EM) material that are electronically controlled with integrated electronics and aim to modify the current distribution over themselves in a deliberate manner to enable unique EM functionalities, including wave absorption, anomalous reflection, polarized reflection, wave splitting, wave focusing, and phase modification. Recent results have revealed that these unique EM functionalities are possible without complex decoding, encoding, and radio frequency (RF) processing operations and the communication system performance can be boosted by exploiting the implicit randomness of wireless propagation \cite{Basar_Access_2019}. However, the fundamental issues remain unsolved within the theoretical and practical understanding as well as modeling of RIS-aided communication systems.

In contrast to current wireless networks, where the environment is out of control of the operators, RISs have enabled the emerging concept of intelligent communication environments, where the environment is turned into an intelligent entity that plays an active role in processing signals and accordingly transferring information. This is a completely new paradigm in wireless communication and has the potential to change the way the communication takes place. The core technology behind this promising concept, RISs, is the metasurfaces, which are the 2D equivalent of metamaterials. Metasurfaces are thin planar artificial structures with sub-wave-length thickness and enable unnatural EM functionalities for the RF, Terahertz, and optical spectrum. It is worth noting that communications through RISs is different compared with other related technologies currently employed in wireless networks, such as relaying, MIMO beamforming, passive reflect-arrays, and backscatter communications, while having the following major distinguishable features:\\
\hspace*{0.5cm}i) RISs are nearly passive, and, ideally, they do not need any dedicated energy source for RF signal processing; \\
\hspace*{0.5cm}ii) RISs do not amplify or introduce noise when reflecting the signals and provide an inherently full-duplex transmission;\\
\hspace*{0.5cm} iii) RISs can be easily deployed, e.g., on the facades of buildings, ceilings of factories, and indoor spaces;\\
\hspace*{0.5cm}  iv) RISs are reconfigurable in order to adapt themselves according to the changes of the wireless environment. 
  
These distinctive characteristics make RIS-assisted communication a unique technology and introduce important communication theoretical as well as system design challenges, some of which will be tackled in this paper.

There has been a growing recent interest in controlling the propagation environment or exploiting its inherently random nature to increase the QoS and/or spectral efficiency. For instance, IM-based \cite{IM_5G,Basar_2017} emerging schemes such as media-based modulation \cite{Khandani1,Basar_2019} and spatial scattering modulation \cite{SSM} use the variations in the signatures of received signals by exploiting reconfigurable antennas or scatterers to transmit additional information bits in rich scattering environments. On the other hand, RISs are smart devices that intentionally control the propagation environment by exploiting reconfigurable reflectors/scatterers to boost the signal quality at the receiver. Although some early attempts have been reported to control the wireless propagation, such as intelligent walls \cite{Subrt_2012,Subrt_2012_2}, spatial microwave modulators \cite{MDR-XX}, 3D reflectors \cite{Xiong_2017}, and coding metamaterials \cite{Cui_2014}, the surge of intelligent communication environments can be mainly attributed to programmable (digitally-controlled) metasurfaces \cite{Yang_2016}, reconfigurable reflect-arrays \cite{Tan_2016,Tan_2018}, software-controlled hypersurfaces \cite{Akyildiz_2018}, and intelligent metasurfaces \cite{MDR-16}. For instance, the intelligent metasurface design of \cite{MDR-16}, enables tuneable anomalous reflection as well as perfect absorption by carefully adjusting the resistance and the capacitance of its unit cells at 5 GHz. 

The concept of communications through intelligent surfaces has received tremendous interest from wireless communication and signal processing communities very recently due to challenging problems it brings in the context of communication, optimization, and probability theories \cite{Di_Renzo_2019,Basar_Access_2019,Wu_2019,DiRenzo_2020}. Particularly, researchers focused on \\
\hspace*{0.5cm}i) maximization of the achievable rate, minimum signal-to-interference-plus noise ratio (SINR), and energy efficiency and minimization of the transmit power by joint optimization of the RIS phases and the transmit beamformer \cite{Huang_2018,Huang_2018_2,Huang_2019,Alouini_2019,Schober_2019,Wu_2018},\\ 
\hspace*{0.5cm}ii) maximization of the received signal-to-noise ratio (SNR) to minimize the symbol error probability \cite{Basar_2019_LIS},\\
\hspace*{0.5cm}iii) efficient channel estimation techniques with passive RIS elements as well as deep learning tools to reduce the training overhead and to reconfigure RISs \cite{Mishra_2019,Taha_2019,He_2019},\\
\hspace*{0.5cm}iv) PHY security solutions by joint optimization of the transmit beamformer and RIS phases \cite{Schober_2019_2,Chen_2019,Cui_2019},\\
\hspace*{0.5cm}v) practical issues such as erroneous reflector phases, realistic phase shifts, and discrete phase shifts \cite{Coon_2019,Abeywickrama_2019,Wu_2018_2}, 
\hspace*{0.5cm}vi) design of NOMA-based systems for downlink transmit power minimization and for the minimum decoding SINR maximization of all users \cite{Fu_2019,Ding_2019}, and\\ 
\hspace*{0.5cm}vii) channel modeling and measurements for different frequency bands and RIS types \cite{Tang_2019,DiRenzo_2020_2,SimRIS1,SimRIS2,SimRIS3,Yildirim_2020}.

Furthermore, the first attempts on combining RISs with space modulation, visible light and free space optical communications, unmanned aerial vehicles, wireless information and power transfer systems, and OFDM systems have been reported in recent times (see \cite{Basar_Access_2019,DiRenzo_2020,Wu_2020} and references therein).

In our paper, we  take a step back and revisit the well-known phenomenons of multipath fading and Doppler effect in mobile communications from the perspective of emerging RISs. Although the potential of RISs has been explored from many aspects as discussed above, to the best of our knowledge,
their potential in terms of Doppler effect mitigation has not been fully understood yet. For this purpose, by following a bottom-up approach from simple networks to more sophisticated ones, we explore the potential of RISs to  eliminate multipath fading effect stemming from Doppler frequency shifts of a mobile receiver for the first time in the literature. Specifically, we prove that the rapid fluctuations in the received signal strength due to the user movement can be effectively eliminated and/or mitigated by utilizing real-time tuneable RISs. We introduce a number of novel and effective methods that provide interesting trade-offs between Doppler effect mitigation and average received signal strength maximization, for the reconfiguration of the available RISs in the system and reveal their potential for future mobile networks.

It is worth noting that the results in this paper are obtained for hypothetical RISs which create specular reflections with a single and very large conducting element. However, the results obtained in this paper can be adapted for practical RISs in which many number of tiny elements on them scatter the incoming signals in all directions, in other words, for RISs with multiplicative path loss terms. Finally, it has been also shown that by using carefully positioned and relatively large RISs, it might be possible to reach the same path gain as that of specular reflection by carefully adjusting the phases of RIS elements \cite{Ellingson}. Nevertheless, exploration of application scenarios in which practical RISs might be effectively used to overcome Doppler and multipath effects is an open problem and this paper aims to shed light in this direction by following a unified signal processing perspective.

The rest of the paper is organized as follows. In Section II, we consider a simple two-path scenario and revisit multipath and Doppler effects. In Section III, we deal with Doppler effect elimination with RISs. Section IV deals with more sophisticated networks with multiple RISs and objects. Finally, in Section V we cover practical issues and in Section VI we conclude the paper.

\section{Revisiting Multipath and Doppler Effects with Simple Case Studies}
In this section, we revisit the Doppler and multipath fading effects caused by the movement of a mobile receiver under a simple propagation scenario (with and without an RIS). We focus our attention to the low-pass equivalent and noise-free received signals while a generalization to pass-band signaling is straightforward from the given low-pass equivalent signals.

\subsection{Multipath Fading Due to User Movement and A Reflector}
\begin{figure}[!t]
	\begin{center}
		\includegraphics[width=1\columnwidth]{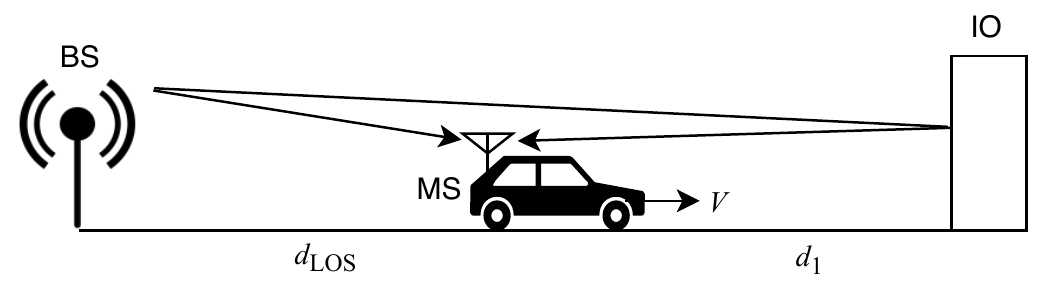}
		\vspace*{-0.6cm}\caption{The basic two-ray propagation model with a mobile receiver and an IO as a reflector.}\vspace*{-0.3cm}
		\label{fig:CaseI}
	\end{center}
\end{figure}

We consider the propagation geometry of Fig. \ref{fig:CaseI} with a base station (BS), a mobile station (MS) that travels along a straight route with a speed of $V$ (in m/s), and an interacting object (IO). In this setup, in addition to the LOS signal stemming from the BS, a second copy of the transmitted signal reflected from the IO arrives at the receiver of the MS. For ease of presentation, we consider a reflection coefficient of unity magnitude and phase $\pi$, that is $R=-1$, for the IO. Here, the reflecting surface is large and smooth enough so that specular reflections occur according to the Snell's law.

\begin{figure}[!t]
	\begin{center}
		\includegraphics[width=1\columnwidth]{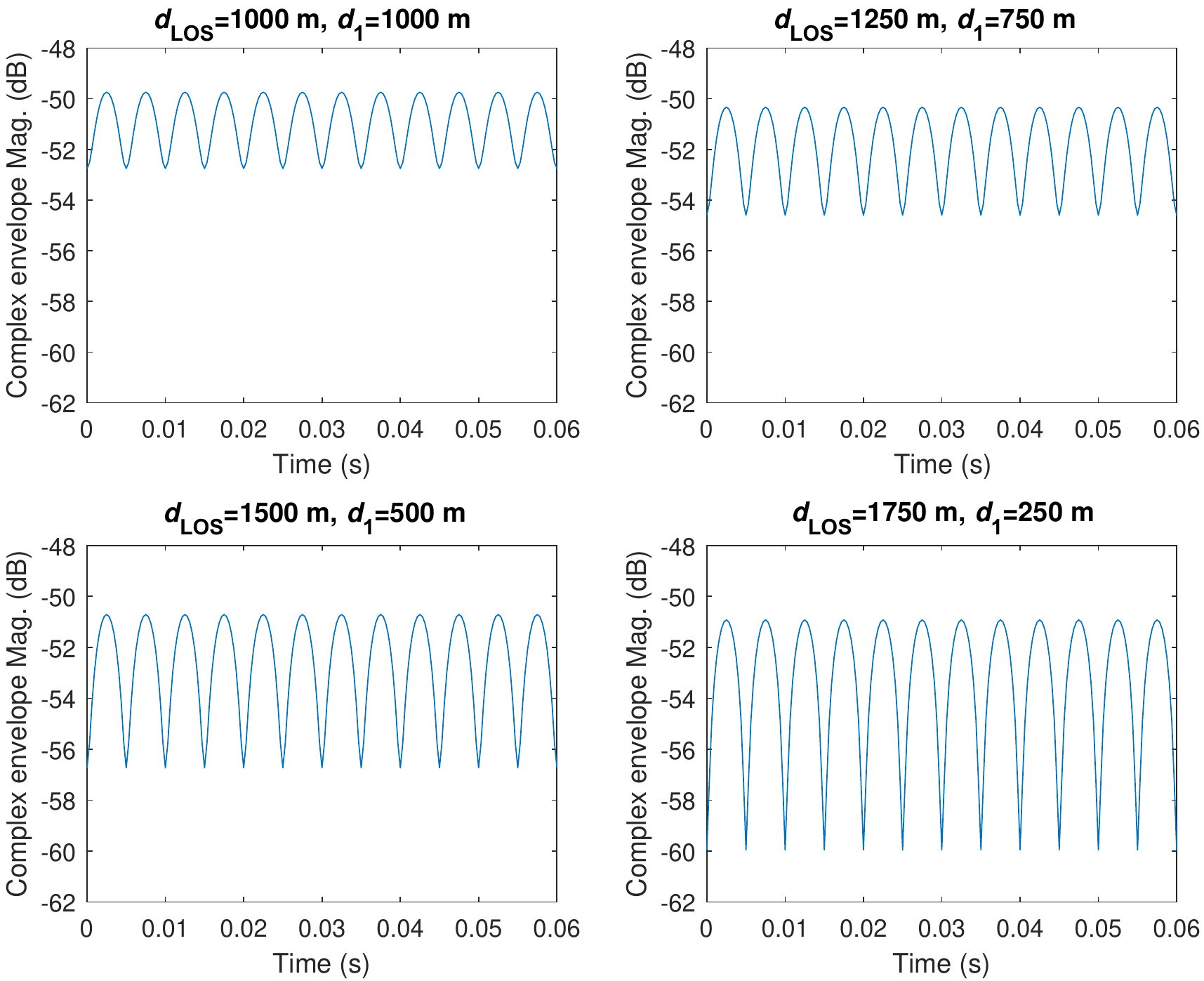}
		\vspace*{-0.6cm}\caption{Variation of the magnitude of the received complex envelope due to MS movement with $V=10$ m/s for varying $d_{\text{LOS}}$ and $d_1$ (observation interval: $0.06$ s, travelled distance: $6\lambda=0.6$ m).}\vspace*{-0.3cm}
		\label{fig:Fig2}
	\end{center}
\end{figure}

\begin{figure}[!t]
	\begin{center}
		\includegraphics[width=1\columnwidth]{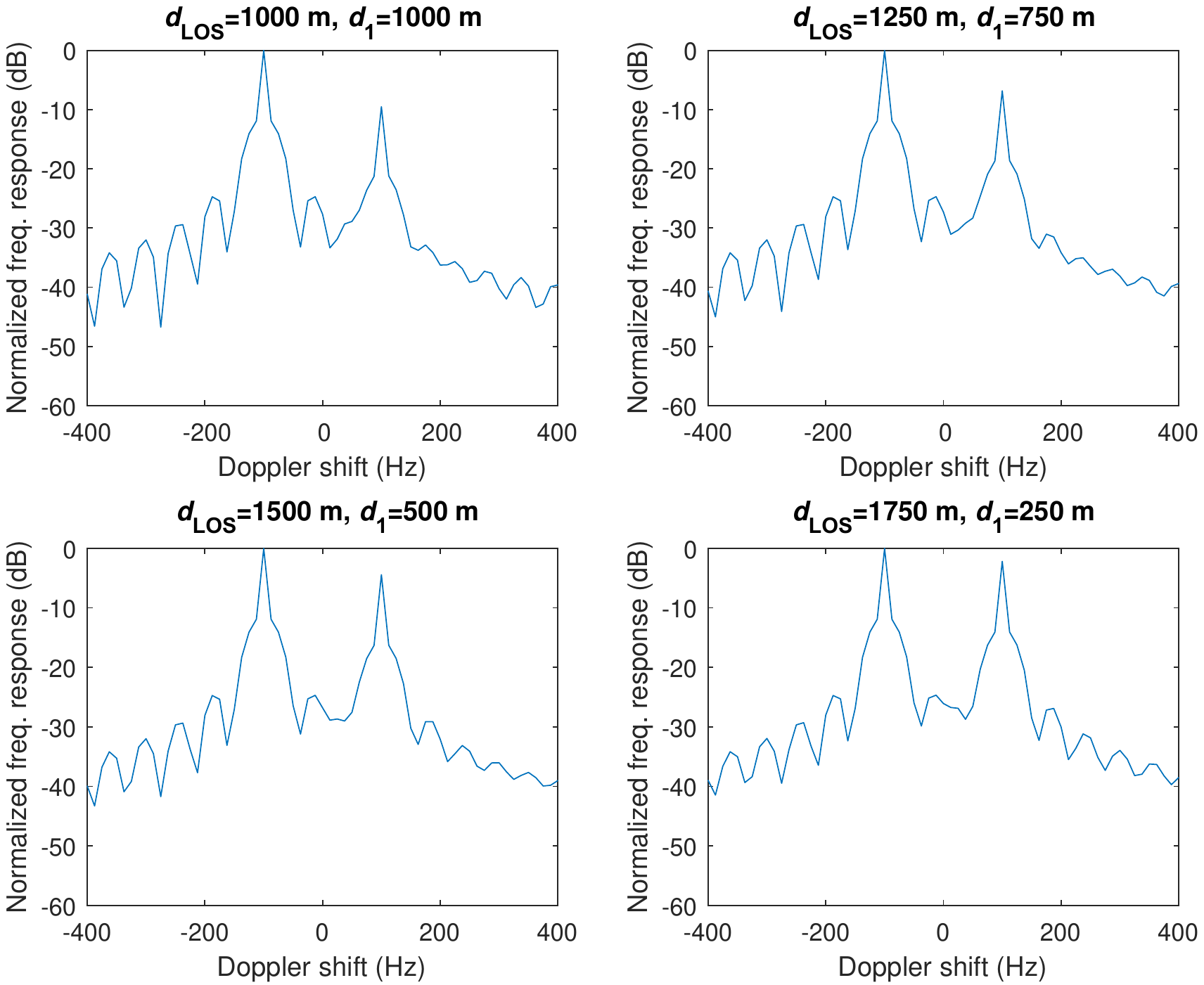}
		\vspace*{-0.6cm}\caption{Doppler spectrum of the received signal for varying $d_{\text{LOS}}$ and $d_1$ (FFT size: $256$, sampling distance: $\lambda/32=0.003125$ m, sampling time: $ \lambda/(32\times V)=0.3125 $ ms).}\vspace*{-0.3cm}
		\label{fig:Fig3}
	\end{center}
\end{figure}

In order to capture the effects of  Doppler and multipath fading in the received signal with respect to time, we assume the transmission of an unmodulated radio frequency (RF) carrier signal $\cos(2\pi f_c t+ \Theta_0)$, where $f_c$ is the carrier frequency and $\Theta_0$ is the initial phase. Using complex baseband representation, we arrive at the low-pass complex equivalent of this signal as $x(t)=\exp(j \Theta_0)$. To illustrate the fade pattern and the Doppler spectrum due to the user movement, we focus on a very short travel distance (a few wavelengths) of the MS, as a result, the received direct and reflected signals have almost constant amplitudes, while being subject to rapidly varying phase terms. At the same time, due to the movement of the MS, Doppler shifts are observed at the received signals. In light of this information, the received complex envelope is obtained as \cite{Fontan}
\begin{equation}\label{eq:1}
r(t)= \frac{\lambda}{4\pi}\left(  \frac{e^{ -j\frac{2\pi d_{\text{LOS}}(t)}{\lambda}}}{d_{\text{LOS}}(t)}  - \frac{e^{ -j\frac{2\pi d_R(t)}{\lambda}}}{d_R(t)}    \right)
\end{equation}
where we dropped the initial carrier phase $\Theta_0$ for clarity. Here $d_{\text{LOS}}(t)$ and $d_R(t)$ respectively stand for the time-varying radio path distance for the BS-MS and the BS-IO-MS links. For the particular setup considered in Fig. \ref{fig:CaseI}, we have $d_{\text{LOS}}(t)=d_{\text{LOS}}+Vt$ and $d_R(t)=d_{\text{LOS}}+2d_1-Vt$, where $d_{\text{LOS}}$ and $d_1$ stand for the initial distances between the BS and the MS and the MS and the IO, respectively. Here, we assume that the BS-MS antenna height difference is sufficiently small so that a horizontal communication link can be considered. The same applies from the signal reflected from the IO. In other words, the radio path length variations for both rays are directly proportional to the MS travel distance variations. However, a generalization is straightforward for signals coming from different angles (see Section IV). As discussed earlier, since we focus into a very short time interval (travel distance of the MS), we may assume that two rays have almost constant amplitudes at the initial and last positions of the MS. Considering this, \eqref{eq:1} simplifies to
\begin{equation}\label{eq:2}
r(t)= \frac{\lambda}{4\pi}\left(  \frac{e^{ -j 2\pi f_D t-j\phi_{\text{LOS}}}}{d_{\text{LOS}}}  - \frac{e^{ j 2\pi f_D t-j\phi_1}}{d_{\text{LOS}}+2d_1}    \right)
\end{equation} 
where $f_D=V/\lambda$ is Doppler shift with respect to the nominal carrier frequency in the passband or with respect to $0$ Hz in low-pass equivalent representation, $\phi_{\text{LOS}}= 2\pi d_{\text{LOS}} / \lambda$, and $\phi_1= 2\pi (d_{\text{LOS}}+2d_1) / \lambda$. The constant (initial) phase terms of $\phi_{\text{LOS}}$ and $\phi_1$ can be readily dropped if they are integer multiples of $2\pi$. In light of this, using the properties of complex exponentials\footnote{For $z_1=r_1e^{j\xi_1}$, $z_2=r_2e^{j\xi_2}$, and $z_3=z_1+z_2=r_3e^{j\xi_3}$, we have $r_3=(r_1^2 + r_2^2 +2r_1r_2\cos(\xi_1-\xi_2))^{1/2}$.}, the magnitude of the complex envelope can be obtained as follows:
\begin{align}\label{eq:3}
&\left| r(t)\right| = \nonumber \\
&   \left( \frac{\lambda}{4\pi}\right) \!\! \left( \frac{1}{d_{\text{LOS}}^2} + \frac{1}{(d_{\text{LOS}}+2d_1)^2} -\frac{2 \cos(4\pi f_D t)}{d_{\text{LOS}}(d_{\text{LOS}}+2d_1)} \right)^{1/2}\!\!\!\!.   
\end{align} 
In Fig. \ref{fig:Fig2}, we plot the magnitude of the complex envelope for an MS travel distance of six wavelengths (corresponding to an observation time of $ 0.06 $ s) considering the following system parameters\footnote{The same simulation parameters (mobile speed, carrier frequency, observation interval, travelled distance, FFT size, sampling distance and time) are used in the following unless specified otherwise.}: $f_c=3 $ GHz, $V=10$ m/s with varying $d_{\text{LOS}}$ and $d_1$ values for a fixed BS-IO total distance of $d_{\text{LOS}}+d_1=2000$ m. As seen from Fig. \ref{fig:Fig2}, due to the destructive and constructive interference of the arriving two signals, the received signal strength fluctuates rapidly (with a frequency of $2f_D$ as evident from \eqref{eq:3}) around a mean value, which is determined by the path loss. This oscillation is also known as the \textit{fade pattern} of the received envelope. It is also worth noting that the variation of the magnitude is more significant for the closer values of $d_{\text{LOS}}(t)$ and $d_R(t)$ (smaller $d_1$). This is also verified by the Doppler spectrum of the received signal given in Fig. \ref{fig:Fig3} for these four cases, which include two sharp components at opposite frequencies, i.e., $-V/\lambda=-100$ Hz (from the LOS path) and $V/\lambda=100$ Hz (from the IO) with different normalized amplitudes due to different travel distances of the two rays.

\subsection{Eliminating Multipath Fading Due to User Movement with an RIS}
In this subsection, we consider the same system model of Subsection II.A (Fig. \ref{fig:CaseI}), however, we assume that a controllable reflection occurs from the IO through an RIS that is mounted on its facade. In this scenario, intelligent reflection is captured by a time-varying and unit-gain reflection coefficient $R(t)=e^{j\theta(t)}$. As a result, the received complex envelope is obtained as
\begin{equation}\label{eq:4}
r(t)= \frac{\lambda}{4\pi}\left(  \frac{e^{ -j 2\pi f_D t}}{d_{\text{LOS}}}  + \frac{e^{ j 2\pi f_D t+j\theta(t)}}{d_{\text{LOS}}+2d_1}    \right).
\end{equation} 
It is obvious that the magnitude of $r(t)$ is maximized when the phases of the direct and reflected signals are aligned, that is, by adjusting the RIS reflection phase as $\theta(t)=-4 \pi f_D t\,\,\,(\mathrm{mod}\,\,\,2\pi)$. It is worth noting that this can be only possible with an RIS that is able to adjust its reflection coefficient dynamically with respect to time (user movement). The practical issues related to this adjustment procedure are discussed in Section V. With the specified value of $\theta(t)$ given above, the complex envelope of the received signal becomes
\begin{equation}\label{eq:5}
r(t)=\frac{\lambda e^{ -j 2\pi f_D t} }{4\pi} \left(\frac{1}{d_{\text{LOS}}} +\frac{1}{d_{\text{LOS}}+2d_1} \right)
\end{equation}
whose magnitude is maximized and remain constant with respect to time during our observation interval and given by
\begin{equation}\label{eq:6}
\left| r(t)\right|_{\max} = \frac{\lambda  }{4\pi} \left(\frac{1}{d_{\text{LOS}}} +\frac{1}{d_{\text{LOS}}+2d_1} \right).
\end{equation}
In light of \eqref{eq:5} and \eqref{eq:6}, we have the following two remarks.

\begin{figure}[!t]
	\begin{center}
		\includegraphics[width=7.5cm,height=5.5cm]{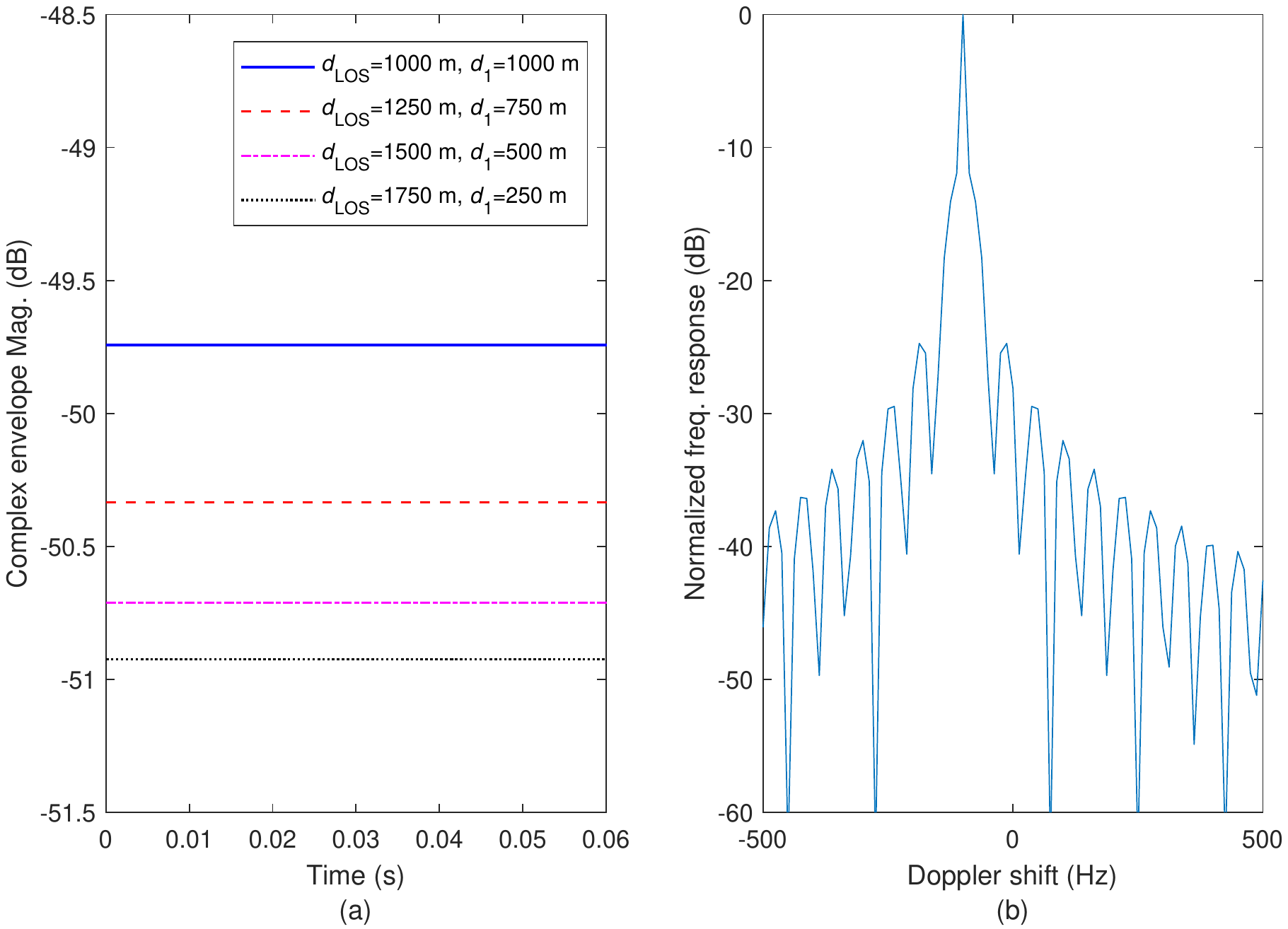}
		\vspace*{-0.3cm}\caption{(a) Maximized magnitude of the received signal in the presence of an RIS, (b) Doppler spectrum of the received signal for all scenarios.}\vspace*{-0.3cm}
		\label{fig:Fig4}
	\end{center}
\end{figure}

\textit{Remark 1}: Time-varying intelligent reflection of the RIS eliminates the multipath fading (rapid fluctuations of the received signal strength) for the scenario of Fig. \ref{fig:CaseI} and enables a constant magnitude for the received complex envelope, which is also shown in Fig. \ref{fig:Fig4}(a). In other words, it is possible to escape from rapid fluctuations in the received signal due to the user movement by utilizing an RIS, which has a time-varying reflection phase.

\textit{Remark 2}: The received signal is still subject to a Doppler shift of $-f_D$ Hz, which is also observed from the Doppler spectrum of Fig. \ref{fig:Fig4}(b). Although the RIS effectively eliminates fade patterns, due to the direct signal received from the BS, which is out of the control of the RIS, it is not possible to eliminate Doppler frequency shifts in this propagation scenario.

\textit{Remark 3}: For the case of a practical RIS in which the incoming signals are scattered with many number of tiny RIS elements in Fig. 1, the received complex envelope is obtained as
\begin{align}\label{eq:RIS_real}
r(t)=  & \frac{\lambda }{4\pi d_{\text{LOS}}}  e^{ -j 2\pi d_{\text{LOS}}(t) / \lambda} \nonumber\\  
&+ \sum_{n=1}^{N} \frac{\lambda^2 G_e}{(4\pi)^2 (d_{\text{LOS}}+d_1) (d_1-Vt)} e^{j (\theta_n(t) -  2\pi d_{R_n}(t)/\lambda) }
\end{align}
where $G_e$ is the element gain, $\theta_n(t)$ is the adjustable phase of the $n$th RIS element and $d_{R_n}(t)$ is the associated path length \cite{SimRIS1}. Here, under the case of far-field, the same path loss is assumed for all RIS elements. By carefully adjusting the RIS phases, that is, for $\theta_n(t)=2\pi (d_{R_n}(t) -   d_{\text{LOS}}(t) ) / \lambda $, the signals coming from the RIS can be aligned to the LOS signal and the magnitude of the complex envelope can be kept constant. Please note that by carefully positioning the RIS and adjusting its size, the magnitude of the complex envelope can be maximized by overcoming the multiplicative path loss in \eqref{eq:RIS_real}. Due to the broad scope of the current paper, detailed investigation of RISs with many scattering elements is left as a future study, while the presented results can be generalized in a systematic way.

\subsection{Increasing Fading and Doppler Effects with An RIS}
So far, we focused our attention on the maximization of the received signal strength for the scenario of Fig. 1, by carefully adjusting the RIS reflection phase in real time. On the contrary, it might be possible to intentionally degrade the received signal strength as well as increase the Doppler spread for an unintended mobile receiver or for an eavesdropper. Based on the received signal model of \eqref{eq:4}, when the received two signals are in-phase, we obtain the maximum magnitude for the received signal as in \eqref{eq:6}. On the other hand, adjusting the RIS reflection phase as $\theta(t)=-4 \pi f_D t+\pi\,\,\,(\mathrm{mod}\,\,\,2\pi)$, we obtain completely out-of-phase two arriving signals, and the resulting minimum complex envelope magnitude becomes
\begin{equation}\label{eq:7}
\left| r(t)\right|_{\min} = \frac{\lambda  }{4\pi} \left(\frac{1}{d_{\text{LOS}}} -\frac{1}{d_{\text{LOS}}+2d_1} \right).
\end{equation}
As seen from \eqref{eq:7}, the degradation in the received signal strength would be more noticeable for smaller $d_1$. However, the magnitude of the complex envelope becomes constant as in \eqref{eq:6}, i.e., no fade patterns are observed. In Fig. \ref{eq:5}(a), we depict the minimized complex envelope magnitudes by intentionally out-phasing the direct and reflected signals for varying $d_{\text{LOS}}$ and $d_1$. Comparing Figs. \ref{eq:4}(a) and \ref{eq:5}(a), we observe up to $9$ dB degradation in magnitude (for $d_{\text{LOS}}=1750$ m and $d_1=250$ m), which corresponds to a power variation of $18$ dB. In other words, it is possible to enable up to $18$ dB variation in the received signal power by deliberately co-phasing and out-phasing the multipath components in the considered setup. It is worth noting that the normalized Doppler spectrum in Fig. \ref{eq:5}(b) is the same as that of Fig. \ref{eq:4}(b).

\begin{figure}[!t]
	\begin{center}
	\includegraphics[width=7.5cm,height=5.5cm]{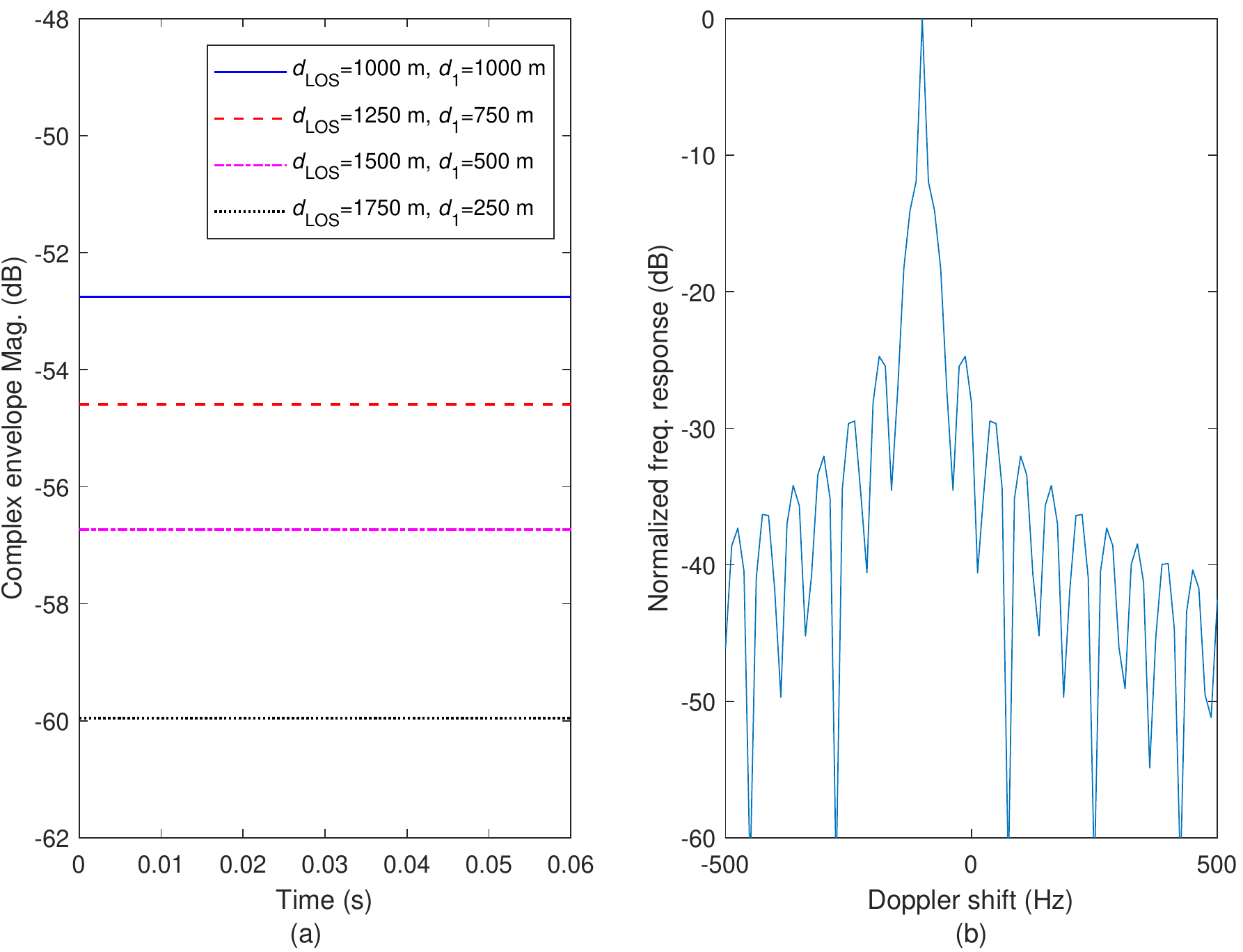}
		\vspace*{-0.3cm}\caption{(a) Minimized magnitude of the received signal in the presence of an RIS, (b) Doppler spectrum of the received signal for all scenarios.}\vspace*{-0.3cm}
		\label{fig:Fig5}
	\end{center}
\end{figure}

\begin{figure}[!t]
	\begin{center}
		\includegraphics[width=1\columnwidth]{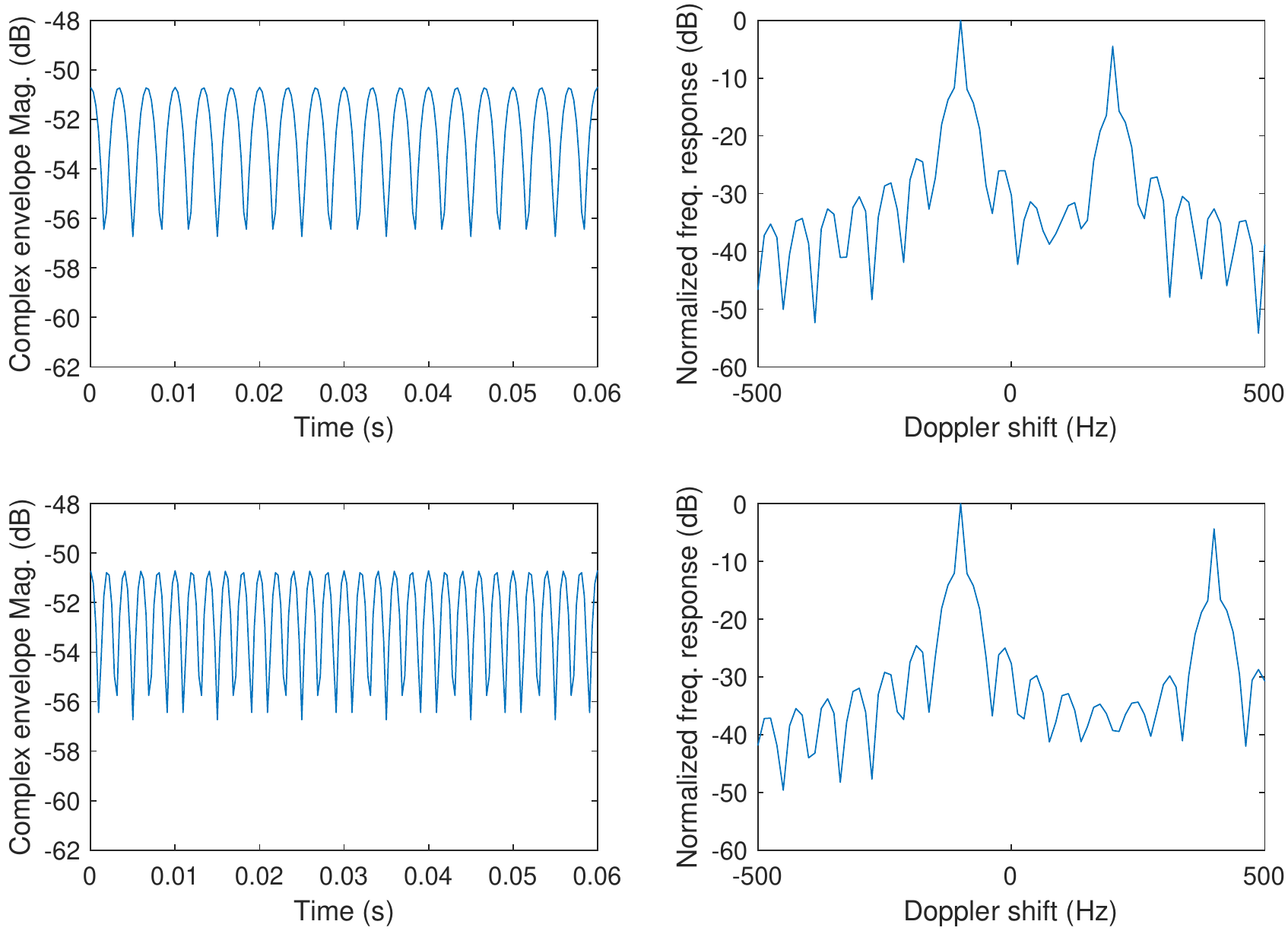}
		\vspace*{-0.3cm}\caption{Time varying magnitude and the Doppler spectrum of the received signal with increased Doppler effect for (top) $\tilde{f}_D=200$ Hz and (bottom) $\tilde{f}_D=400$ Hz.}\vspace*{-0.3cm}
		\label{fig:Fig6}
	\end{center}
\end{figure}

As another extreme application of an RIS, the Doppler spread can be increased by intentionally increasing the Doppler shift of the reflected signal by $\theta(t)=2\pi (\tilde{f}_D-f_D) t\,\,\,(\mathrm{mod}\,\,\,2\pi) $, where $\tilde{f}_D$ is the desired Doppler shift for the reflected signal. Here, a maximum desired Doppler shift of $0.5 f_s$ Hz can be observed in simulation, where $f_s $ is the sampling frequency for the continuous-wave signal. In Fig. \ref{eq:6}, we present the magnitude of the complex envelope as well as the Doppler spectrum for the case of $d_{\text{LOS}}=1500$ m and $d_1=500$ m by carefully adjusting the reflection phase to increase the Doppler spread (reduce the coherence time) by $\tilde{f}_D=200$ Hz and $400$ Hz. As seen from Fig. \ref{eq:6}, an RIS can create new components in the Doppler spectrum, which results in a faster fade pattern for the complex envelope.

Going one step further, we consider the concept of random phase shifts by the RIS, in which the reflection phase is selected at random between $0$ and $2 \pi$ in each time interval. We illustrate the magnitude of the complex envelope and the Doppler spectrum in Fig. \ref{eq:7} for the case random reflection phases, where the reflection phase is selected at random in each sampling time for $d_{\text{LOS}}=1500$ m and $d_1=500$ m. As seen from Figs. 7(a) and (b), although the effect in Doppler spectrum is not very significant, it would be possible to obtain a very fast fade pattern in time. Specifically, around $5$ dB magnitude variations are observed within a sampling distance of $\lambda/32$ m. It would be possible to obtain an ultra-fast fade pattern by alternating the reflection phase between $\theta(t)=-4 \pi f_D t+\pi\,\,\,(\mathrm{mod}\,\,\,2\pi)$ and $\theta(t)=-4 \pi f_D t\,\,\,(\mathrm{mod}\,\,\,2\pi)$ in each time interval and this is left for interested readers.

\begin{figure}[!t]
	\begin{center}
		\includegraphics[width=7.5cm,height=5.5cm]{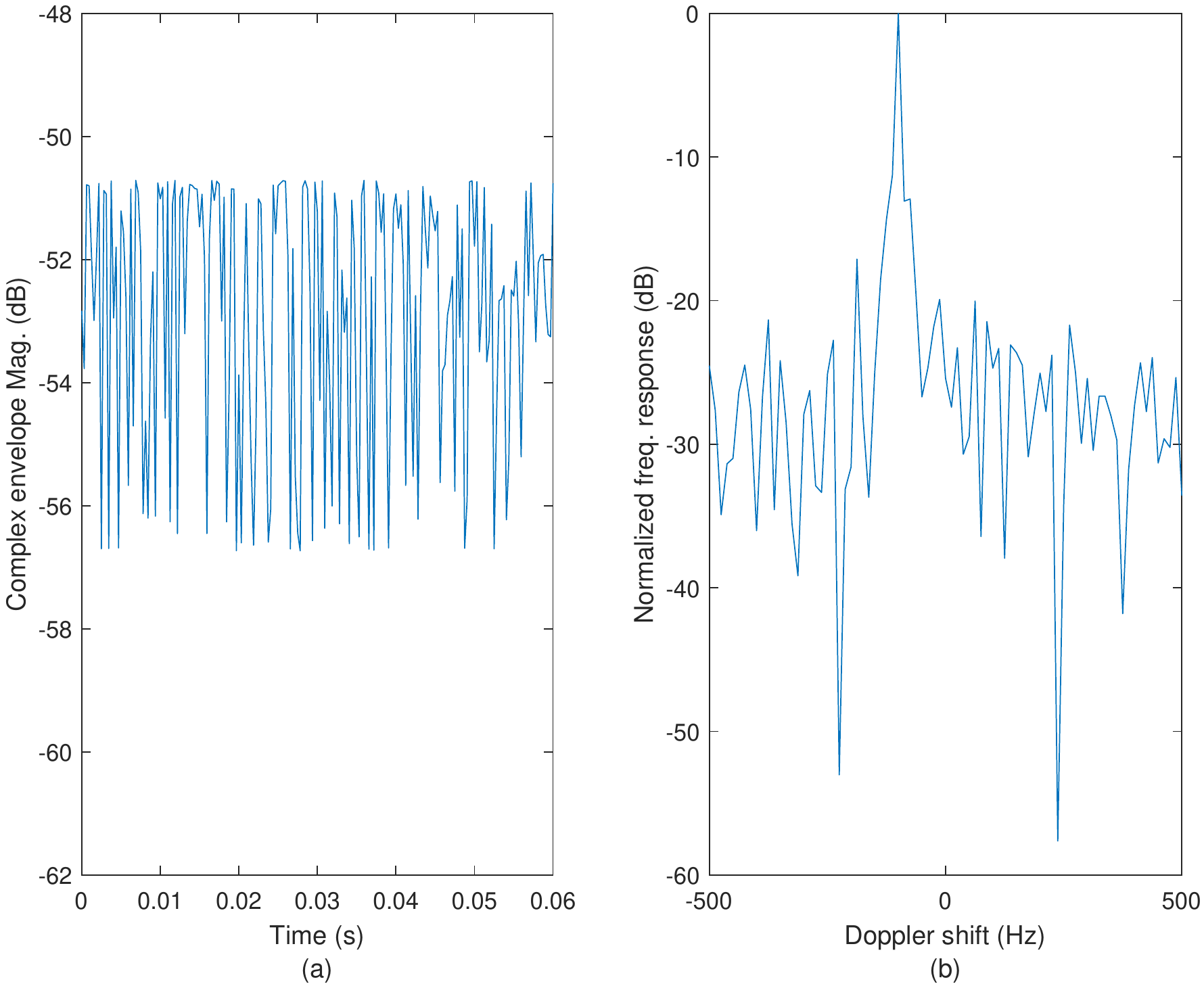}
		\vspace*{-0.3cm}\caption{(a) Magnitude of the received signal in the presence of an RIS with random phases, (b) Doppler spectrum of the received signal.}\vspace*{-0.3cm}
		\label{fig:Fig7}
	\end{center}
\end{figure}

\section{Eliminating Doppler Effects Through Intelligent Reflection}
In this section, we focus on a simple scenario in which the direct link is blocked by an obstacle while the communication between the BS and the MS is established through a reflection from an IO as shown in Fig. 8. We consider the same assumptions of Section II and investigate the Doppler effect on the received signal in the following two cases.

\subsection{NLOS Transmission without An RIS}   
Under the assumption of specular reflections from the IO with a reflection coefficient of $R=-1$, the received signal can be expressed as
\begin{equation}\label{eq:8}
r(t)= -\frac{\lambda}{4\pi}  \frac{e^{ -j\frac{2\pi d_R(t)}{\lambda}}}{d_R(t)}  
\end{equation}
where $d_R(t)=d_{\text{LOS}}+2d_1-Vt$ is the time-varying radio path distance for a MS moving with a speed of $V$ m/s. Ignoring the constant phase terms and assuming a very short travel distance, the received signal can be expressed as
\begin{equation}\label{eq:9}
r(t)= - \frac{\lambda e^{ j 2\pi f_D t}}{4\pi(d_{\text{LOS}}+2d_1)}.    
\end{equation} 
As seen from \eqref{eq:9}, since only a single reflection occurs without a LOS signal and other multipath components, the received signal magnitude does not exhibit a fade pattern, that is, fixed with respect to time and given by $ \left|r(t) \right|=\lambda/(4\pi (d_{\text{LOS}}+d_1)) $. However, the received signal is still subject to a Doppler frequency shift of $f_D$ Hz, which is evident from \eqref{eq:9}, due to the movement of the MS. 

\begin{figure}[!t]
	\begin{center}
		\includegraphics[width=1\columnwidth]{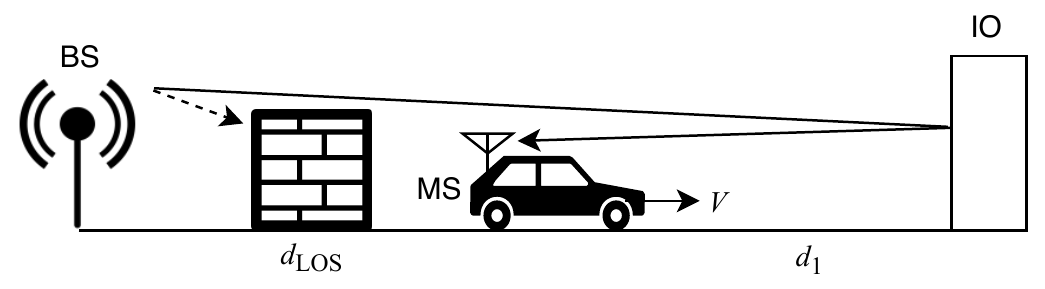}
		\vspace*{-0.6cm}\caption{Communications through an IO with a blocked LOS path.}\vspace*{-0.3cm}
		\label{fig:CaseII}
	\end{center}
\end{figure}

\subsection{NLOS Transmission with An RIS}  
Here, we focus on the scenario of Fig. \ref{fig:CaseII} while assuming that the IO is equipped with an RIS that is able to provide adjustable phase shifts, that is, $R(t)=e^{j \theta(t)}$, as in Section II. In this case, the received signal can be expressed as
\begin{equation}\label{eq:10}
r(t)=  \frac{\lambda e^{ j 2\pi f_D t + j\theta(t)}}{4\pi(d_{\text{LOS}}+2d_1)}.    
\end{equation} 
As seen from \eqref{eq:10}, the magnitude of the received signal is independent from the reflection phase and the same as the previous case (without an RIS). However, it might be possible to completely eliminate the Doppler effect by adjusting the RIS reflection phase as $\theta(t)=-2 \pi f_D t\,\,\,(\mathrm{mod}\,\,\,2\pi)$. We give the following remark.

\textit{Remark 3}: When there is no direct transmission between the BS and the MS over which the RIS has no control, intelligent reflection allows one to completely eliminate the Doppler effect, by carefully compensating the Doppler phase shifts through the RIS. 

In Fig. 9, we show the Doppler spectrum of the received signal with and without an RIS for $d_{\text{LOS}}=d_1=1000$ m. As seen from Fig. 9, the Doppler effect is eliminated ($0$ Hz) by adjusting the RIS reflections accordingly.

As discussed earlier, it is not possible the modify the magnitude of the complex envelope with an RIS for the scenario of Fig. \ref{fig:CaseII}, however, as done in Section II, the Doppler spread can be enhanced by $\theta(t)=2\pi (\tilde{f}_D-f_D) t\,\,\,(\mathrm{mod}\,\,\,2\pi) $, where $\tilde{f}_D$ is the desired Doppler frequency. The observation of the resulting spectrum is straightforward and left for the interested readers.

\begin{figure}[!t]
	\begin{center}
		\includegraphics[width=7.5cm,height=5.5cm]{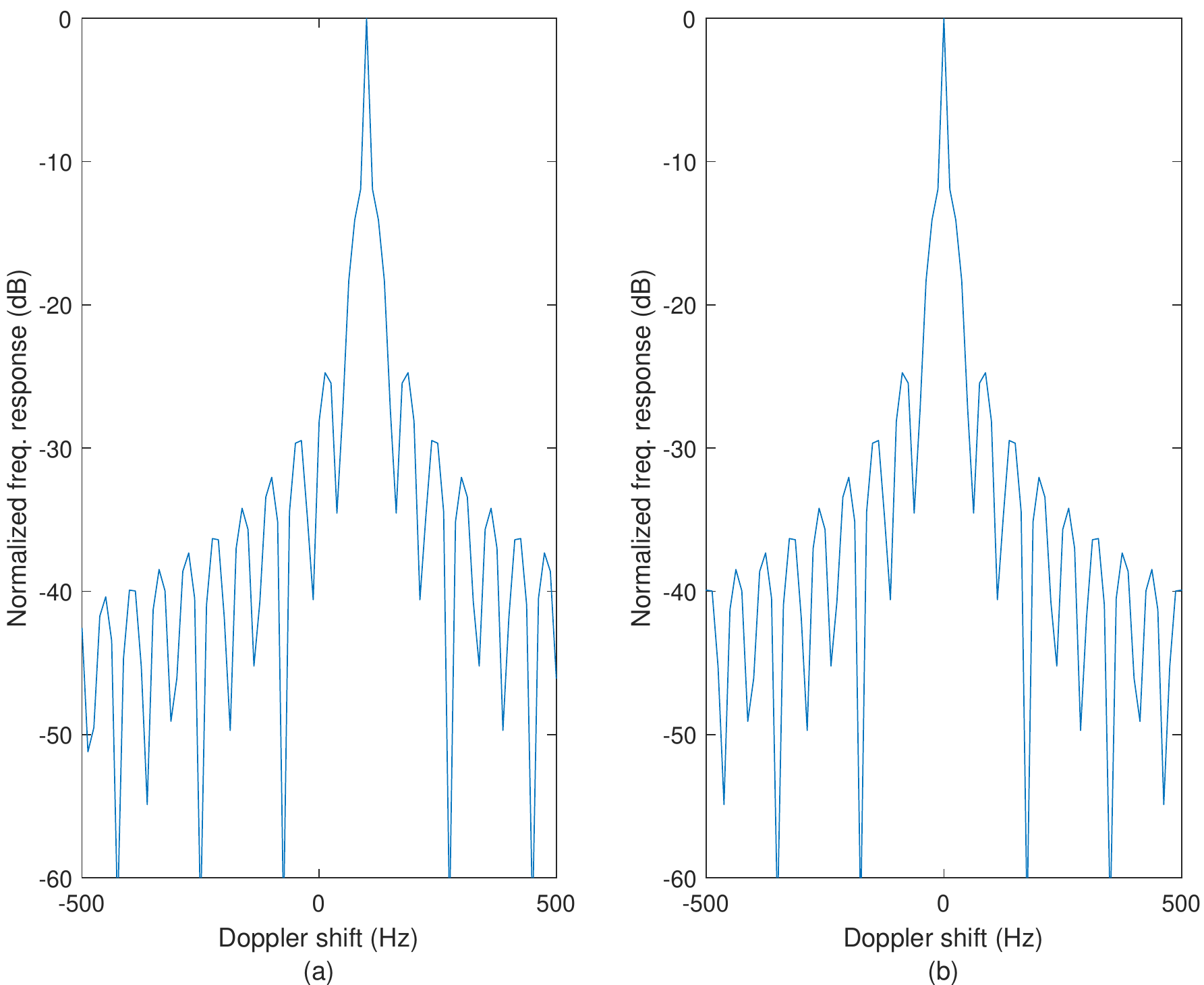}
		\vspace*{-0.3cm}\caption{Doppler spectrum of the received signal for the scenario of Fig. \ref{fig:CaseII}: (a) without an RIS, (b) with an RIS.}\vspace*{-0.3cm}
		\label{fig:Fig9}
	\end{center}
\end{figure}

\section{Doppler and Multipath Fading Effects: Case Studies with Multiple Reflectors}
In this section, we extend our system models and analyses in Sections II and III into propagation scenarios with multiple IOs with and without intelligent reflection capabilities. We follow a bottom-up approach starting with two IOs and illustrate the fading/Doppler effect mitigation capabilities of RISs. We also propose a  number of effective and novel methods with different functionalities.

\subsection{Direct Signal and Two Reflected Signals without any RISs}

\begin{figure}[!t]
	\begin{center}
		\includegraphics[width=1\columnwidth]{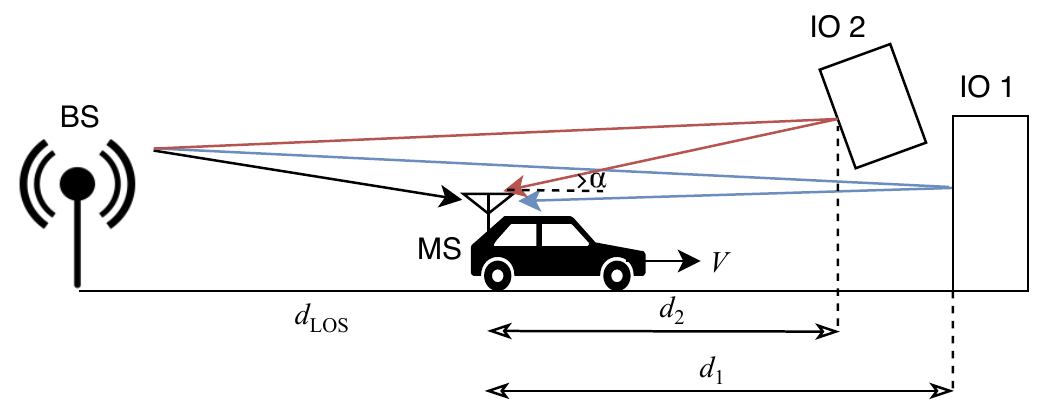}
		\vspace*{-0.6cm}\caption{Propagation scenario with two IOs.}\vspace*{-0.3cm}
		\label{fig:CaseIII}
	\end{center}
\end{figure}

In this subsection, by extending our model given in Section II, we consider the propagation scenario of Fig. \ref{fig:CaseIII} with two IOs. Here, in order to spice up our analyses, we assume that while the BS-MS and BS-IO 1-MS links are parallel to the ground, the reflected signal from IO 2 arrives to the MS with an angle of $\alpha$ with respect to the MS route. In this scenario, the initial (horizontal) distances between the BS and the MS, the MS and IO 1, and the MS and IO 2 are shown by $d_{\text{LOS}}$, $d_1$, and $d_2$, respectively. Using a similar analysis as in Section II, under the assumption of unit gain reflection coefficients for both IOs, that is $R_1=R_2=-1$, the time-varying received complex envelope can be expressed as 
\begin{equation}\label{eq:11}
r(t)= \frac{\lambda}{4\pi}\left(  \frac{e^{ -j 2\pi f_D t}}{d_{\text{LOS}}}  - \frac{e^{ j 2\pi f_D t}}{d_{\text{LOS}}+2d_1} - \frac{e^{ j 2\pi f_D(\cos \alpha) t-j\phi_2}}{\tilde{d_2} }   \right)
\end{equation} 
where $\tilde{d_2}=\sqrt{d_2^2 \tan^2 \alpha + (d_{\text{LOS}}+d_2 )^2} + d_2/(\cos\alpha)$ is the initial radio path distance for the reflected signal of IO 2, which is obtained after simple trigonometric operations, and  $\phi_2=2\pi \tilde{d_2}/\lambda$ is a fixed phase term. Here, we assume that the variations in terms of the large-scale path loss due to the movement of MS are almost negligible (as in Fig. \ref{fig:CaseI}) and the rays from IO 2 remain parallel for all points of the mobile route, which corresponds to radio path distance decrements of $Vt\cos \alpha$, with respect to time, for these rays. It is worth noting that parallel ray assumption is approximately true for short route lengths \cite{Goldsmith}. As seen from \eqref{eq:11}, the received signal has three Doppler components: $-f_D$ Hz, $f_D$ Hz, and $f_D\cos\alpha$ Hz due to the rays coming from the BS, IO 1, and IO 2, respectively.   

\begin{figure}[!t]
	\begin{center}
		\includegraphics[width=7.5cm,height=5.5cm]{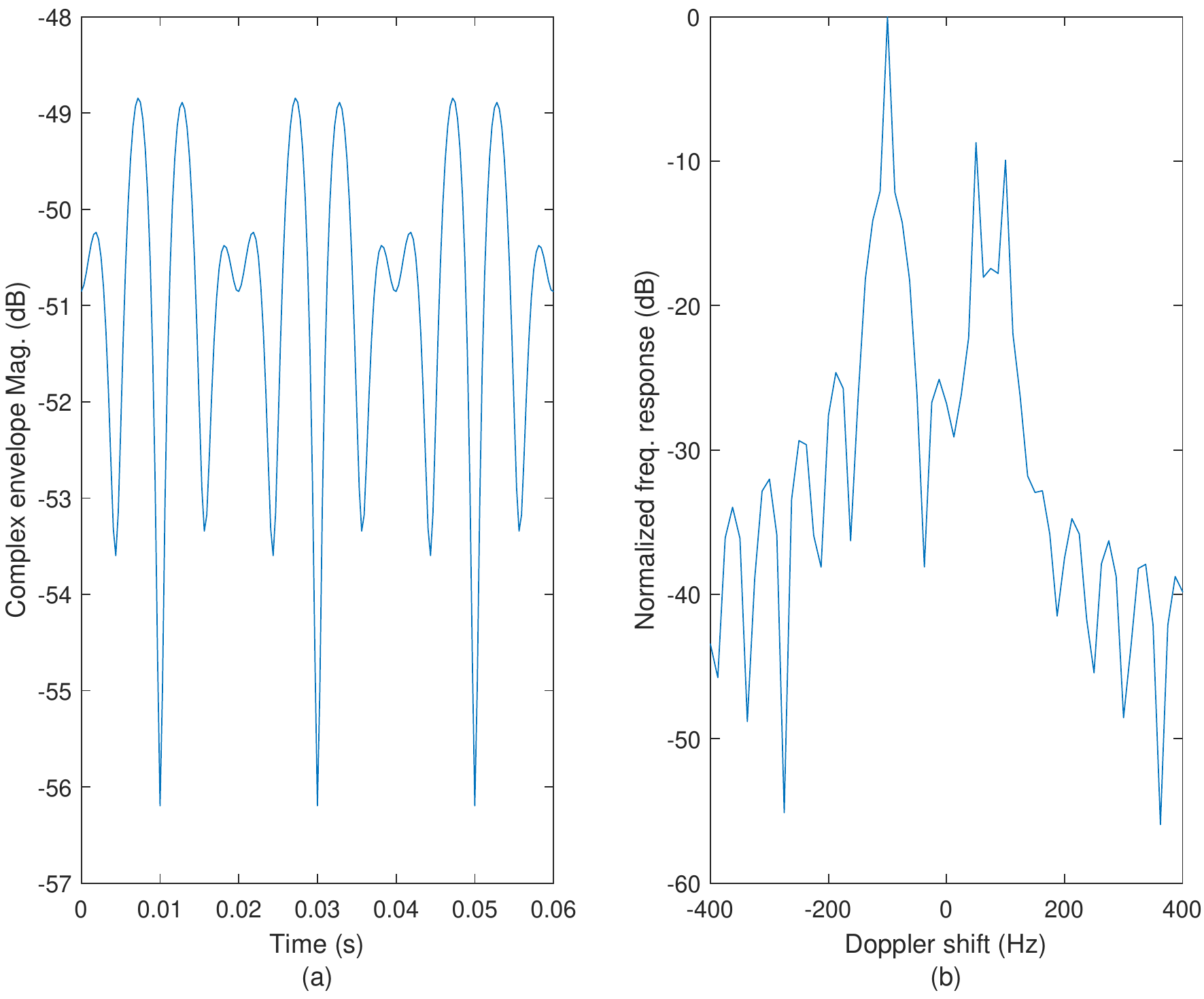}
		\vspace*{-0.3cm}\caption{The scenario of Fig. \ref{fig:CaseIII} without an RIS: (a) Complex envelope magnitude, (b) Doppler spectrum of the received signal.}\vspace*{-0.3cm}
		\label{fig:Fig11}
	\end{center}
\end{figure}

In Fig. \ref{fig:Fig11}, we show the magnitude of the complex envelope as well as the Doppler spectrum for the case of $\alpha=60^\circ$, $d_{\text{LOS}}=d_1=1000$ m, and $d_2=500$ m. As seen from Fig. \ref{fig:Fig11}, due to constructive and destructive interference of the direct and two reflected signals with different Doppler frequency shifts ($-100$ Hz, $100$ Hz, and $50$ Hz), the magnitude of the complex envelope exhibits a more hostile and faster fading pattern compared to the simpler scenario of Fig. \ref{fig:CaseI} (see Fig. \ref{fig:Fig2}, top-left subplot).

\subsection{Direct Signal and Two Reflected Signals with One or Two RISs}

In this subsection, we again focus on the scenario of Fig. \ref{fig:CaseIII}, however, under the assumption of one or two RISs that are attached to the existing IOs. Although being more challenging in terms of system optimization and analysis, we focus on the case of a single RIS first, then extend our analysis into the case of two RISs.

\subsubsection{One RIS}
\begin{figure}[!t]
	\begin{center}
		\includegraphics[width=1\columnwidth]{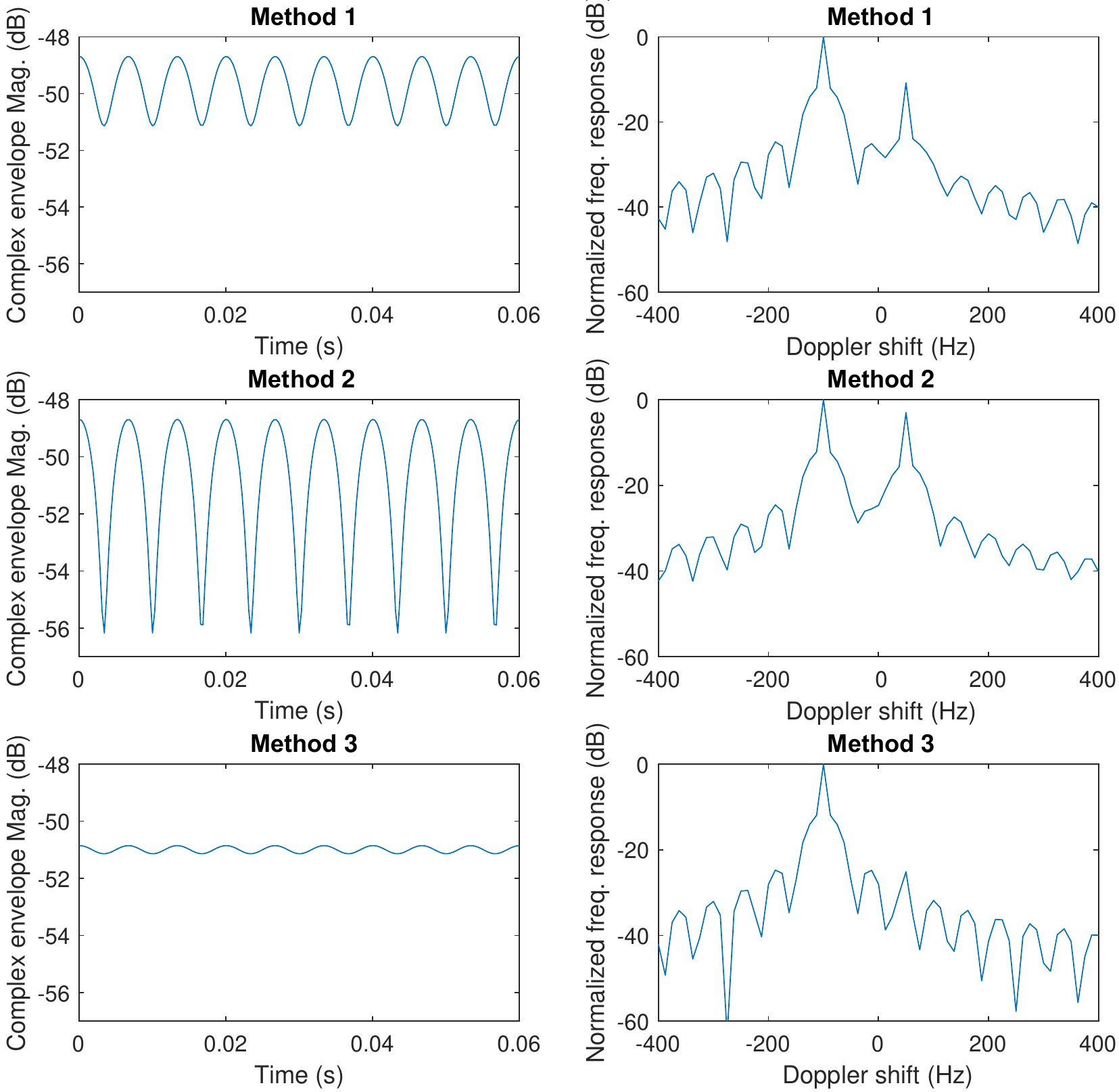}
		\vspace*{-0.3cm}\caption{Magnitude and Doppler spectrum of the received signal with an RIS for scenario of Fig. 10 under three different phase selection methods.}\vspace*{-0.3cm}
		\label{fig:Fig12}
	\end{center}
\end{figure}

Let us assume that we have a single RIS that is mounted on the facade of IO 1 for the scenario of Fig. \ref{fig:CaseIII}. For this case, the received complex envelope can be expressed as 
\begin{align}\label{eq:12}
&r(t)= \frac{\lambda}{4\pi}\left(  \frac{e^{ -j 2\pi f_D t}}{d_{\text{LOS}}}  + \frac{e^{ j 2\pi f_D t+j\theta_1(t)}}{d_{\text{LOS}}+2d_1}\right.  \nonumber \\
&\hspace*{3cm}\left. - \frac{e^{ j 2\pi f_D(\cos \alpha) t-j\phi_2}}{\tilde{d_2} }   \right).
\end{align} 
Here, we assumed that the intelligent reflection from IO 1 is characterized by $\theta_1(t)$. We investigate the following three methods for the adjustment of $\theta_1(t)$, where the corresponding complex envelope magnitudes and Doppler spectrums are shown in Fig. \ref{fig:Fig12} for $\alpha=60^\circ$, $d_{\text{LOS}}=d_1=1000$ m, and $d_2=500$ m:
\begin{itemize}
	\item Method 1: $ \theta_1(t) = -4 \pi f_D t\,\,\,(\mathrm{mod}\,\,\,2\pi)  $
	\item Method 2: $ \theta_1(t) = 2 \pi f_D t(\cos \alpha -1)-\phi_2+\pi\,\,\,(\mathrm{mod}\,\,\,2\pi)  $
	\item Method 3: $ \theta_1(t) = 2 \pi f_D t(\cos \alpha -1)-\phi_2\,\,\,(\mathrm{mod}\,\,\,2\pi)  $
\end{itemize}
In the first method, we intuitively align the reflected signal from the RIS to the LOS signal. As seen from Fig. \ref{fig:Fig12}, although this adjustment eliminates the $100$ Hz component in the spectrum and reduces the Doppler spread compared to the case without RIS (Fig. \ref{fig:Fig11}), we still observe two components in the spectrum and a noticeable fade pattern for the received signal due to uncontrollable reflection through IO 2. It is worth noting that this might be the preferred option to obtain a high time average for the complex envelope magnitude with the price of a high Doppler spread (faster time variation). 

In the second method, we align the reflected signal from the RIS to the one from IO 2, however, this worsens the situation by increasing the relative power of the $50$ Hz component in the Doppler spectrum. As seen from Fig. \ref{fig:Fig12}, a more severe fade pattern is observed for Method 2 due to destructive interference of the reflected signals to the LOS signal. This would be a preferred option in case of an eavesdropper to degrade its signal quality.

In the third method,  we follow a clever approach and instead of aligning our RIS-assisted reflected signal to the existing two signals, we target to eliminate the uncontrollable reflection from IO 2 by out-phasing the reflected two signals. This results a remarkable improvement in both Doppler spectrum and the received complex envelope by almost mitigating the fade pattern. In other words, the RIS scarifies itself in Method 3 to eliminate the uncontrollable reflection from IO 2, which significantly reduces the multipath effect, while a minor variation is still observed due to different radio path lengths of these two signals. More specifically, for the selection of $\theta_1(t)$ in Method 3, we obtain
\begin{align}\label{eq:13}
&r(t)= \frac{\lambda}{4\pi}\left(  \frac{e^{ -j 2\pi f_D t}}{d_{\text{LOS}}} \right.  \nonumber \\
&\left. \hspace*{1cm}+ e^{ j 2\pi f_D(\cos \alpha) t-j\phi_2} \left( \frac{1}{d_{\text{LOS}}+2d_1} - \frac{1}{\tilde{d_2}} \right) \right) 
\end{align} 
which contains two components. However, the Doppler spread can be remarkably reduced when the radio path distances of the signals reflected from IO 1 and 2, i.e., $d_{\text{LOS}}+2d_1$ and $\tilde{d_2}$, are close to each other. For instance, for the considered system parameters of $d_{\text{LOS}}$, $d_1$, $d_2$, and $\alpha$ in Fig. \ref{fig:Fig12}, we have $ \frac{1}{d_{\text{LOS}}} \gg \left( \frac{1}{d_{\text{LOS}}+2d_1} - \frac{1}{\tilde{d_2}}\right) $, which results almost a single-tone received signal $r(t) \approx  \frac{\lambda}{4\pi}\left(  \frac{e^{ -j 2\pi f_D t}}{d_{\text{LOS}}} \right)$. This is also evident from the Doppler spectrum of the received signal for Method 3. However, Method 3 cannot guarantee the highest complex envelope magnitude, which is also observed from Fig. \ref{fig:Fig12}.

\begin{figure}[!t]
	\begin{center}
		\includegraphics[width=1\columnwidth]{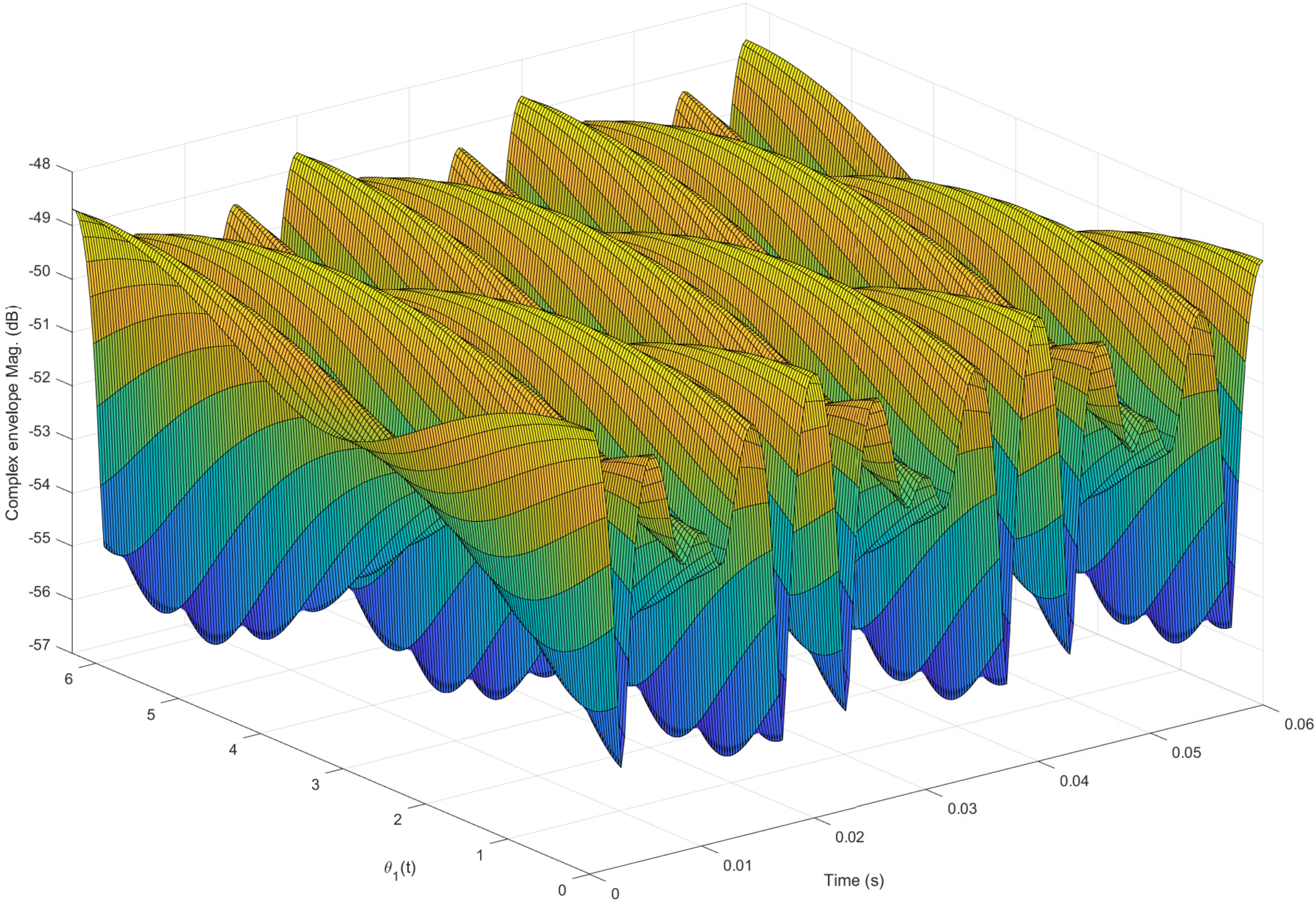}
		\vspace*{-0.3cm}\caption{3D illustration of the variation of the complex envelope magnitude with respect to time for all possible RIS reflection angles (scenario of Fig. \ref{fig:CaseIII}).}\vspace*{-0.3cm}
		\label{fig:Fig13}
	\end{center}
\end{figure}

To gain further insights, in Fig. \ref{fig:Fig13}, we plot the 3D magnitude of the complex envelope with respect to time and varying $\theta_1(t)$ values between $0$ and $2\pi$. As seen from Fig. \ref{fig:Fig13}, due to constructive and destructive interference of multipath components (particularly due to the interference of the signal reflected from IO 2), the complex envelope exhibits several deep fades. We also observe that it is not feasible to fix the complex envelope magnitude to its maximum value ($-48.69$ dB for this specific setup) as in the case of single reflection since the incoming three signals cannot be fully aligned at all times. Finally, we note that performing an exhaustive search for the determination of the optimum reflection phase that maximizes $\left| r(t)\right| $ for each time sample might be possible with different system parameters, however, this does not fit within the scope of this study, which explores effective solutions for the RIS configuration. We also verify from Fig. \ref{fig:Fig13} that Method 1 achieves approximately the maximum magnitude for the complex envelope in the considered experiment. In light of our discussion above, we give the following remark:

\textit{Remark 4}: For the case of two reflections with a single RIS in Fig. \ref{fig:CaseIII}, the heuristic choice to maximize the magnitude of the complex envelope is to align the reflected signal to the stronger component, that is, the LOS signal (Method 1) under normal circumstances. While this ensures a very high magnitude for the complex envelope, we still observe a fade pattern in time domain. On the other hand, the RIS can be reversely aligned to the reflected signal from the plain IO (Method 3) to reduce the Doppler spread at the price of a slight degradation in the magnitude of the complex envelope.

\begin{figure}[!t]
	\begin{center}
		\includegraphics[width=7.5cm,height=5.5cm]{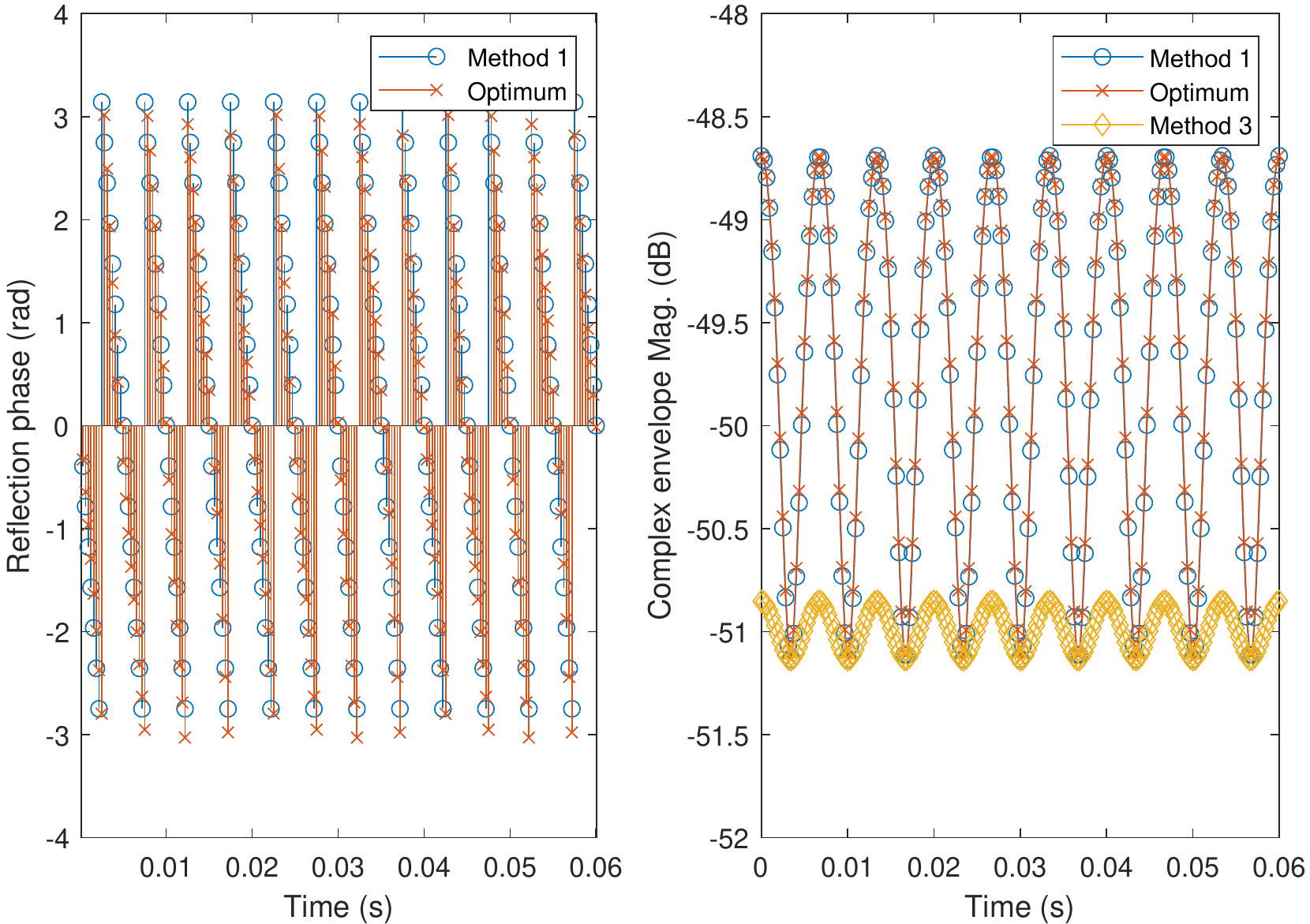}
		\vspace*{-0.3cm}\caption{Comparison of reflection phases and complex envelope magnitudes for Method 1 and the optimum method.}\vspace*{-0.3cm}
		\label{fig:Fig14}
	\end{center}
\end{figure}

\textit{Remark 5}: For the setup of Fig. \ref{fig:CaseIII}, the optimal reflection phase that maximizes the magnitude of the complex envelope is given by
\begin{equation}\label{eq:14}
\theta_1(t)= \frac{\pi}{2}(1-\mathrm{sgn}(A))- \tan^{-1}(-B/A)
\end{equation}  
where $\mathrm{sgn}(\cdot)$ is the sign function and
\begin{align}\label{eq:15}
A=& \frac{1}{d_{\text{LOS}}} \cos(4\pi f_D t) - \frac{1}{\tilde{d_2}}\cos(2 \pi f_D (1-\cos\alpha)t + \phi_2) \nonumber \\
B=& \frac{-1}{d_{\text{LOS}}} \sin(4\pi f_D t) + \frac{1}{\tilde{d_2}}\sin(2 \pi f_D (1-\cos\alpha)t + \phi_2). 
\end{align}
The proof of \eqref{eq:14} is given in Appendix. In Fig. \ref{fig:Fig14}, we compare the reflection phases as well as magnitudes of the complex envelope for Method 1 and the optimum method for the same system parameters. As seen from Fig. \ref{fig:Fig14}, Method 1 provides a very close phase behavior compared to the optimal one due to the stronger LOS path and a very minor degradation can be observed in the magnitude of the complex envelope. Nevertheless, the optimal reflection phase in \eqref{eq:14} is valid for all possible system parameters in Fig. \ref{fig:CaseIII} and guarantees the maximum complex envelope magnitude at all times. For reference, magnitude values are also shown in the same figure for Method 3. As seen from Fig. \ref{fig:Fig14}, Method 3 reduces the severity of the fade pattern (Doppler spread) while ensuring the same minimum magnitude at the price of a lower time average for the complex envelope.

\subsubsection{Two RISs}
Under the assumption of two RISs attached to the existing two IOs in the system of Fig. \ref{fig:CaseIII}, the received complex envelope is obtained as
\begin{align}\label{eq:16}
&r(t)= \frac{\lambda}{4\pi}\left(  \frac{e^{ -j 2\pi f_D t}}{d_{\text{LOS}}}  + \frac{e^{ j 2\pi f_D t+j\theta_1(t)}}{d_{\text{LOS}}+2d_1}\right.  \nonumber \\
&\hspace*{3cm}\left. + \frac{e^{ j 2\pi f_D(\cos \alpha) t-j\phi_2+ j\theta_2(t)}}{\tilde{d_2} }   \right)
\end{align}
where the time-varying and intelligent reflection characteristics of RIS 1 and 2 are captured by $\theta_1(t)$ and $\theta_2(t)$, respectively. Here, compared to the previous case, we have more freedom with two controllable reflections and the magnitude of the received signal can be maximized (and the Doppler spread can be minimized) by readily aligning the reflected signals to the LOS signal. This can be done by setting  $\theta_1(t)=-4 \pi f_D t\,\,\,(\mathrm{mod}\,\,\,2\pi)$ and $\theta_2(t) =- 2 \pi f_D t (1+ \cos \alpha) + \phi_2\,\,\,(\mathrm{mod}\,\,\,2\pi)$, which results
\begin{equation}\label{eq:17}
r(t)=\frac{\lambda e^{ -j 2\pi f_D t}}{4\pi}  \left( \frac{1}{d_{\text{LOS}}} + \frac{1}{d_{\text{LOS}} + 2d_1} + \frac{1}{\tilde{d_2}}\right). 
\end{equation}
Similar to the case with single intelligent reflection (Subsection II.B), we obtain a constant-amplitude complex envelope and a minimized Doppler spread (with a single component at $-f_D$ Hz) due to the clever co-phasing of the multipath components. Interested readers may easily obtain the magnitude and the Doppler spectrum of the complex envelope to verify our findings. 

\subsection{Two RISs without a LOS path}
Finally, we extend our analysis for the case of non-LOS transmission through two RISs, which yields
\begin{align}\label{eq:18}
r(t)= \frac{\lambda}{4\pi}\left(  \frac{e^{ j 2\pi f_D t+j\theta_1(t)}}{d_{\text{LOS}}+2d_1} + \frac{e^{ j 2\pi f_D(\cos \alpha) t-j\phi_2+ j\theta_2(t)}}{\tilde{d_2} }   \right).
\end{align}
Similar to the case in Section III, by carefully adjusting the phases of two RISs, the Doppler effect can be totally eliminated due to the nonexistence of the LOS signal, which is out of control of the RISs. It is evident that this can be done by $\theta_1(t)=-2 \pi f_D t\,\,\,(\mathrm{mod}\,\,\,2\pi)$ and $\theta_2(t)=-2 \pi f_D(\cos\alpha) t+\phi_2\,\,\,(\mathrm{mod}\,\,\,2\pi)$.

\subsection{The General Case with Multiple IOs and the Direct Signal}
\begin{figure}[!t]
	\begin{center}
		\includegraphics[width=1\columnwidth]{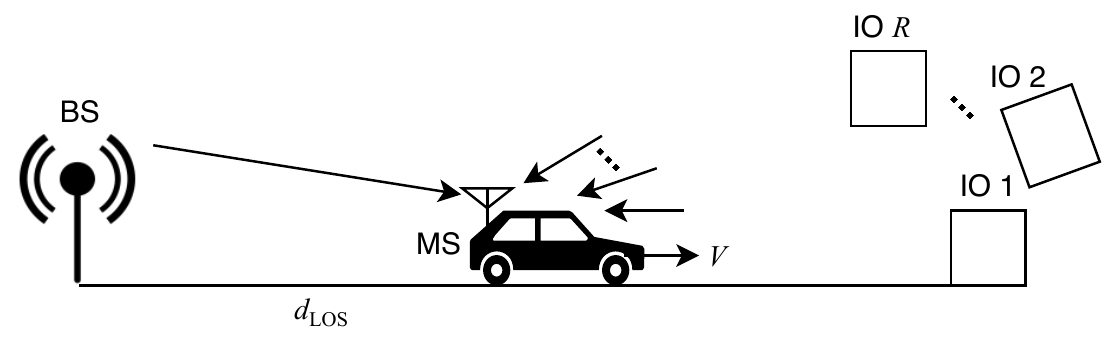}
		\vspace*{-0.6cm}\caption{The general case of multiple  IOs with $N$ RISs and $M$ plain IOs ($R=N+M$).}\vspace*{-0.3cm}
		\label{fig:CaseIV}
	\end{center}
\end{figure}

Against this background, in this subsection, we extend our analyses for the general case of Fig. \ref{fig:CaseIV}, which consists of a total of $R$ IOs. Here, we assume that $N$ of them are coated with RISs, while the remaining $M=R-N$ ones are plain IOs, which create uncontrollable specular reflections towards the MS. In this scenario ($N$ RISs and $M$ plain IOs), the received complex envelope is given by
\begin{align}\label{eq:19}
&r(t)= \frac{\lambda}{4\pi} \left( \frac{e^{ -j 2\pi f_D t}} {d_{\text{LOS}}}+ \sum_{i=1}^{N} \frac{e^{j 2\pi f_{R,i}t-j\psi_i+j\theta_i(t)}}{\tilde{d}_{R,i}} \right. \nonumber \\
&\hspace{2cm}\left. -\sum_{k=1}^{M} \frac{e^{j 2\pi f_{I,k}t-j\phi_k}}{\tilde{d}_{I,k}}\right). 
\end{align}

Here, we assume that all rays stemming from IOs remain parallel during the movement of the MS for a short period of time, which is a valid assumption, and without loss of generality, we consider a reflection coefficient of $-1$ for the plain IOs. Additionally, the corresponding terms in \eqref{eq:19} are defined as follows:
\begin{itemize}
	\itemsep0em 
	\item $f_{R,i}$: Doppler shift for the $i$th RIS
	\item $f_{I,k}$: Doppler shift for the $k$th plain IO
	\item $\psi_{i}$: Constant phase shift for the $i$th RIS
	\item $\phi_{k}$: Constant phase shift for the $k$th plain IO
	\item $\tilde{d}_{R,i}$: Initial radio path distance for the $i$th RIS
	\item $\tilde{d}_{I,k}$: Initial radio path distance for the $k$th plain IO
	\item $\theta_i(t)$: Adjustable phase shift of the $i$th RIS
\end{itemize}
Here, the Doppler shifts of the RISs and plain IOs are not only dependent on the speed of the MS, but also on their relative positions with respect to the MS, i.e., angles of arrival for the incoming signals: $f_{R,i}=f_D \cos \alpha_i $ and $f_{I,k}=f_D \cos \beta_k $, where $\alpha_i$ and $\beta_k$ are the angles of arrival for the reflected signals of $i$th RIS and $k$th plain IO, respectively. In this generalized scenario, we focus on the following two setups:

\subsubsection{Setup I $(N \le M)$}
In this setup, we have more number of uncontrollable reflectors (plain IOs) than RISs. Consequently, we extend our methods in Subsection IV.B and target either directly aligning $N$ RISs to the LOS path (to improve the received signal strength) or  eliminating the reflections stemming from $N$ out of $M$ plain IOs (to reduce the Doppler spread). While the alignment of the reflected signals to the LOS signal is straightforward (Method 1), the assignment of $N$ RISs to corresponding IOs in real-time appears as an interesting design problem. For this purpose, we consider a brute-force search algorithm to determine the most effective set of IOs to be targeted by RISs (Methods 2 \& 3). More specifically, $N$ out of $M$ IOs can be selected in $C(M,N)$ different ways, where $C(\cdot,\cdot)$ is the binomial coefficient. Since these $N$ RISs can be assigned to $N$ plain IOs in $N!$ ways, we obtain a total of $P(M,N)=C(M,N) N!$ possibilities (permutations) for the assignment of $N$ RISs to $M$ IOs. Our methodology has been summarized below:
\begin{itemize}
	\item Method 1: We align the existing $ N $ RISs to the LOS path by adjusting their reflection phases as $\theta_i(t) = - 2\pi f_{R,i}t + \psi_i -2\pi f_D t\,\,\,(\mathrm{mod}\,\,\,2\pi) $ for $i=1,2,\ldots,N$.
	\item Method 2: For the $i$th RIS minimizing the effect of the reflection stemming from the $k$th IO, i.e., $i$th RIS out-phased with the $k$th plain IO, we have the following reflection phase: $\theta_i(t) =  - 2\pi f_{R,i}t + \psi_i + 2\pi f_{I,k}t -\phi_k  \,\,\,(\mathrm{mod}\,\,\,2\pi)$ for $i=1,2,\ldots,N$ and $k=1,2,\ldots,M$. Considering these given reflection phases, for each time instant, we search for all possible $N$-permutations of $M$ plain IOs to maximize the absolute value of the complex envelope. Then, the permutation of IOs that maximizes the  complex envelope magnitude is selected. This method requires a search over $P(M,N)$ permutations in each time instant, in return, has a higher complexity than the first one. Specifically, let us denote the $n$th permutation (the set of IOs) by $\mathcal{P}_n=\left\lbrace \mathcal{P}_n^1,\mathcal{P}_n^2\ldots,\mathcal{P}_n^N \right\rbrace  $ for $n=1,2,\ldots,P(M,N)$. For a given time instant $t=t_0$, considering all permutations, we construct the possible the set of RIS phases as $\theta_i(t_0) =  - 2\pi f_{R,i}t_0 + \psi_i + 2\pi f_{I,\mathcal{P}_n^i}t_0 -\phi_{\mathcal{P}_n^i}  \,\,\,(\mathrm{mod}\,\,\,2\pi)$ for $i=1,2,\ldots,N$ and the corresponding estimate of the received signal sample $r_n(t_0)$ is obtained from \eqref{eq:19} for the $n$th permutation. Finally, the optimum permutation is obtained as $\hat{n}=\arg \max_n \left| r_n(t_0) \right| $. Then, the optimal set of plain IOs to be targeted by RISs are determined as $\mathcal{P}_{\hat{n}}$ and the RIS reflection phases are adjusted accordingly: $\hat{\theta}_i(t_0) =  - 2\pi f_{R,i}t_0 + \psi_i + 2\pi f_{I,\mathcal{P}_{\hat{n}}^i}t_0 -\phi_{\mathcal{P}_{\hat{n}}^i}  \,\,\,(\mathrm{mod}\,\,\,2\pi)$ for $i=1,2,\ldots,N$. These procedures are repeated for all time instants. Obviously, this strategy requires the knowledge of all Doppler phases at a central processing unit, estimation of the received complex envelope samples, and a dynamic control of all RISs. 
	\item Method 3: This method uses the same exhaustive search approach of Method 2, however, instead of maximizing the the absolute value of the complex envelope, we try to minimize the variation of it with respect to time by assigning the RISs to IOs with this purpose. Specifically, for a given time instant $t=t_0$, the optimal permutation $\mathcal{P}_{\hat{n}}$ is obtained as $\hat{n}= \arg \min _n \big| \left|  r_n(t_0)\right|  - \left| r(t_{-1}) \right|  \big| $, where $r(t_{-1})$ is the sample of the received signal at the previous time instant, while at $t=0$, we determine the optimal permutation as in Method 2. This method directly targets to eliminate fade patterns of the complex envelope instead of focusing on the maximization of the received signal strength by aligning (co-phasing) RISs with certain IOs. In other words, Method 3 eliminates the variations in the received signal stemming from different Doppler shifts of the incoming signals.  
\end{itemize}

\subsubsection{Setup II $(N> M)$}
In this setup, we have more number of RISs than the plain IOs, and consequently, have much more freedom in the system design. Here, we consider the same three methods discussed above (Setup I) for the adjustment of RIS reflection phases, however, slight modifications are performed for Methods 2 and 3 due to fewer number of plain IOs in this setup. In Method 1, we align the existing RISs to the LOS path as in Setup I. To reduce the Doppler spread by Method 2, we search for all possible $M$-permutations of RISs to target plain IOs, i.e., a total of $ P(N,M) $ permutations are considered. More specifically, at each time instant, we consider all possible RIS permutations to eliminate the reflections from $M$ plain IOs, while the remaining $N-M$ RISs are aligned to the LOS path. The permutation of RISs that maximizes the absolute value of the sample of the received signal is selected. On the other hand, Method 3 aims to minimize the variations in $r(t)$ by assigning $M$ RISs to $M$ plain IOs, while also aligning the remaining $N-M$ RISs to the LOS path. Our methodology has been summarized as follows:

\begin{itemize}
	\item Method 1: The same as Method 1 for Setup I.
	\item Method 2: Let us denote the $n$th permutation (the set of RISs) by $\mathcal{R}_n=\left\lbrace \mathcal{R}_n^1,\mathcal{R}_n^2\ldots,\mathcal{R}_n^M \right\rbrace  $  and the set of RISs that are not included in the $n$th permutation by $\mathcal{S}_n=\left\lbrace \mathcal{S}_n^1,\mathcal{S}_n^2\ldots,\mathcal{S}_n^{N-M} \right\rbrace  $, i.e., $\mathcal{P}_n \cup \mathcal{S}_n = \left\lbrace 1,2,\ldots,N \right\rbrace $ for $n=1,2,\ldots,P(N,M)$. For a given time instant $t=t_0$, considering all permutations, we construct the possible the set of RIS phases to eliminate IO reflections as $\theta_{\mathcal{R}_n^i}(t_0) =  - 2\pi f_{R,\mathcal{R}_n^i}t_0 + \psi_{\mathcal{R}_n^i} + 2\pi f_{I,i}t_0 -\phi_{i}  \,\,\,(\mathrm{mod}\,\,\,2\pi)$ for $i=1,2,\ldots,M$, while aligning the remaining $N-M$ RISs to the LOS path as follows: $\theta_{\mathcal{S}_n^i}(t_0) = - 2\pi f_{R,\mathcal{S}_n^i}t_0 + \psi_{\mathcal{S}_n^i} -2\pi f_D t_0\,\,\,(\mathrm{mod}\,\,\,2\pi) $ for $i=1,2,\ldots,N-M$. Then, the corresponding estimate of the received signal sample $r_n(t_0)$ is obtained from \eqref{eq:19} for the $n$th permutation. Finally, the optimum permutation is obtained as $\hat{n}=\arg \max_n \left| r_n(t_0) \right| $. Then, the optimal set of RISs to be paired with IOs and aligned to the LOS path are determined as $\mathcal{R}_{\hat{n}}$ and $\mathcal{S}_{\hat{n}}$, respectively, and the RIS reflection phases are adjusted accordingly: $\hat{\theta}_{\mathcal{R}_{\hat{n}}^i}(t_0) =  - 2\pi f_{R,\mathcal{R}_{\hat{n}}^i}t_0 + \psi_{\mathcal{R}_n^i} + 2\pi f_{I,i}t_0 -\phi_{i}  \,\,\,(\mathrm{mod}\,\,\,2\pi)$ for $i=1,2,\ldots,M$ and $\hat{\theta}_{\mathcal{S}_{\hat{n}}^i}(t_0) =\\ - 2\pi f_{R,\mathcal{S}_{\hat{n}}^i}t_0 + \psi_{\mathcal{S}_{\hat{n}}^i} -2\pi f_D t_0\,\,\,(\mathrm{mod}\,\,\,2\pi) $ for $i=M+1,M+2,\ldots,N$. The above procedures are repeated for all time samples.
	\item Method 3: This method follows the same procedures as that of Method 2, except the determination of the optimum permutation. This is performed by $\hat{n}= \arg \min _n \big| \left|  r_n(t_0)\right|  - \left| r(t_{-1}) \right|  \big| $ considering the current (estimated corresponding to the $n$th permutation) and previously received signal samples of $r_n(t_0)$ and $r(t_{-1})$.
\end{itemize}

\begin{figure}[!t]
	\begin{center}
		\includegraphics[width=0.75\columnwidth]{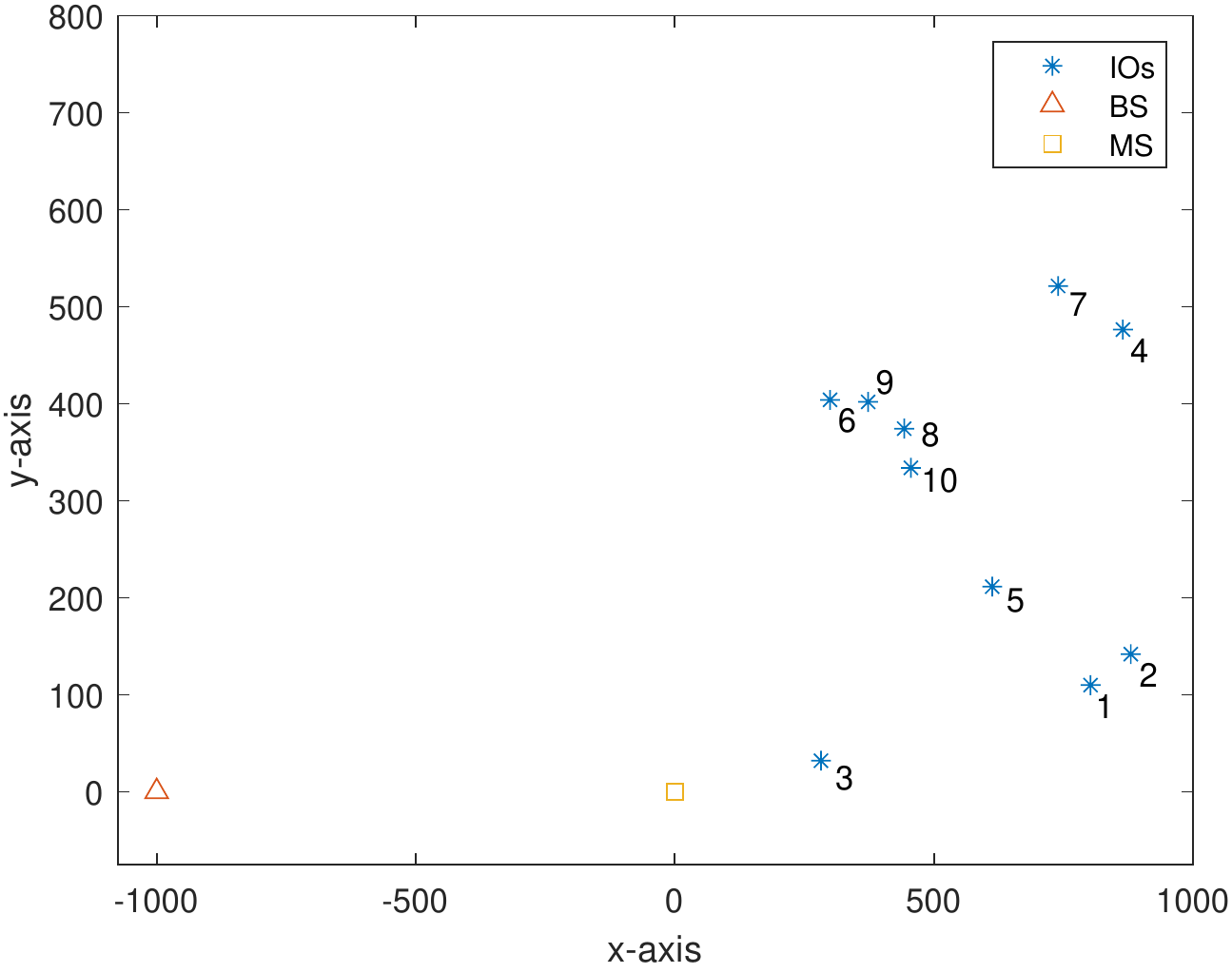}
		\vspace*{-0.3cm}\caption{Considered simulation geometry with multiple IOs (the first $N$ of them are assumed to have RISs).}\vspace*{-0.3cm}
		\label{fig:Fig16}
	\end{center}
\end{figure} 

To illustrate the potential of our methods, we consider the 2D geometry of Fig. \ref{fig:Fig16} in our computer simulations, where the MS and the BS are located at $(0,0)$ and $(-1000,0)$ in terms of their $(x,y)$-coordinates, respectively. We assume that $R=10$ IOs are uniformly distributed in a predefined rectangular area at the right hand side of the origin. We again consider a mobile speed of $V=10$ m/s with $f_c=3$ GHz and a sampling time of $\lambda/32$, but use the following new simulation parameters: a travel distance of $30\lambda=3$ m and an FFT size of $1024$.

\begin{figure}[!t]
	\begin{center}
		\includegraphics[width=1\columnwidth]{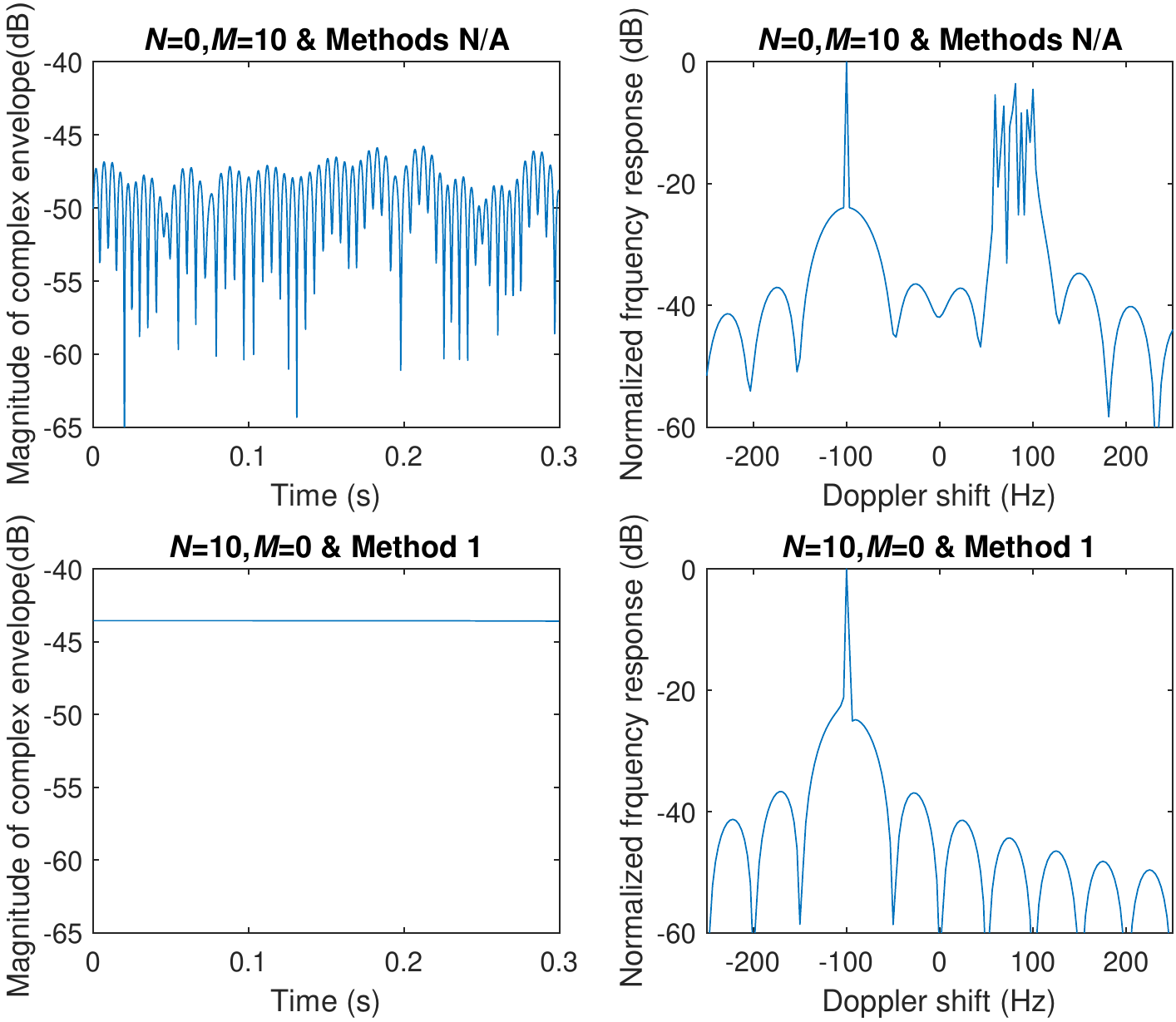}
		\vspace*{-0.3cm}\caption{Complex envelope and Doppler spectrum  for two extreme cases under the scenario of Fig. 15: (top) $N=0,M=10$ (10 plain IOs without any RISs) and (bottom) $N=10,M=0$ (10 RISs without any plain IOs). }\vspace*{-0.3cm}
		\label{fig:Fig17}
	\end{center}
\end{figure} 

In Fig. \ref{fig:Fig17}, we investigate two extreme cases: $N=0,M=10$ and $N=10,M=0$. For the case of $N=0,M=10$, i.e., the case without any RISs, we observe a Doppler spectrum consisting of many components and in return, a severe deep fading pattern in the time domain. On the contrary, for the case of $N=10,M=0$, in which all IOs in the system are equipped with RISs, we have a full control of the propagation environment by applying Method 1 (aligning the reflected signals from all RISs to the LOS path) and observe a constant magnitude for the complex envelope as in Subsections II.B and IV.B.2. Here, we may readily state that the case of  $N=10,M=0$ with Method 1 provides the maximum magnitude for the complex envelope and can be considered as a benchmark for all setups/methods with $M>0$.

\begin{figure}[!t]
	\begin{center}
		\includegraphics[width=1\columnwidth]{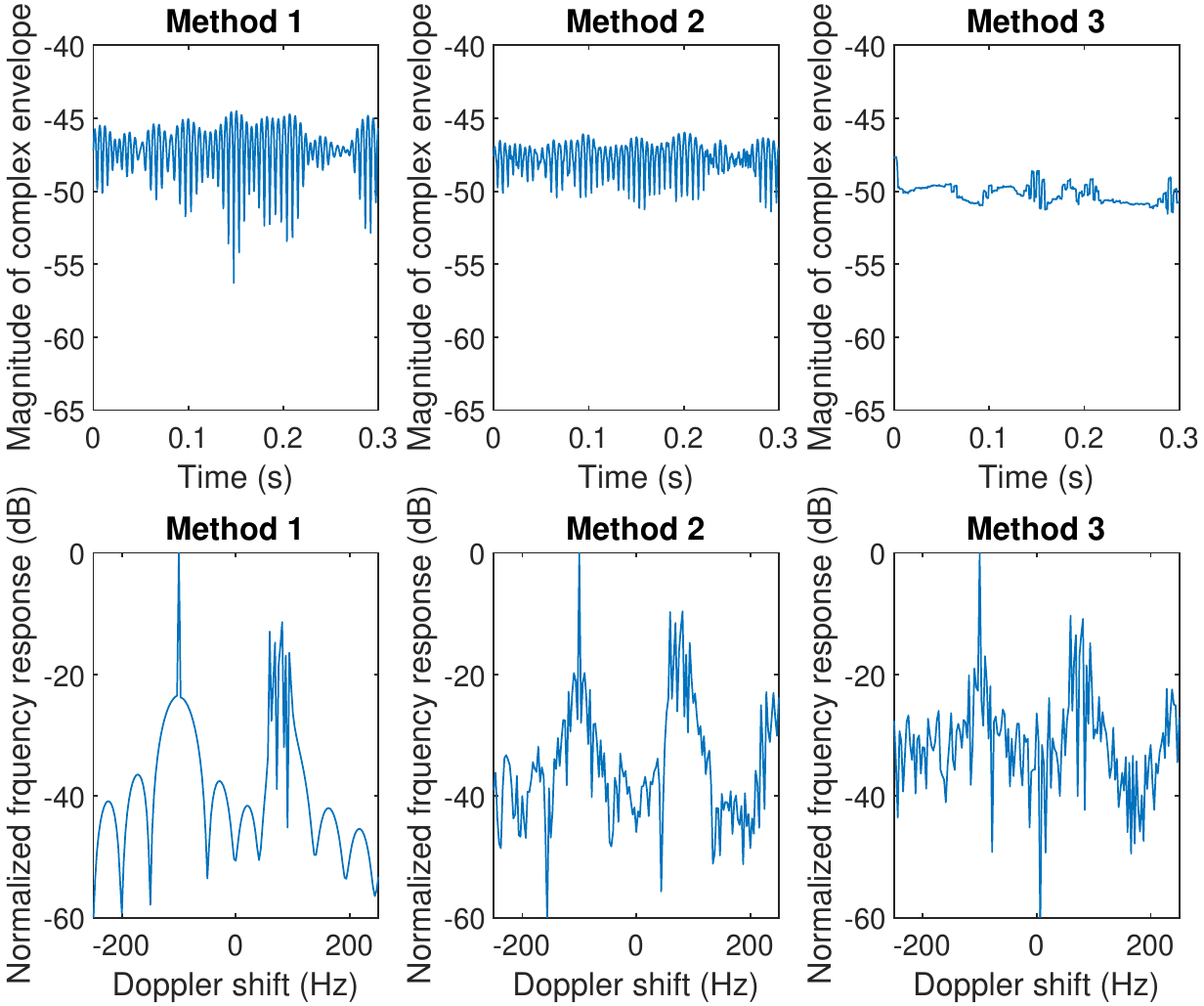}
		\vspace*{-0.4cm}\caption{Complex envelope magnitude and Doppler spectrum for the general case with 10 IOs and $N=3,M=7$ (Setup I).}\vspace*{-0.3cm}
		\label{fig:Fig18}
	\end{center}
\end{figure}

\begin{figure}[!t]
	\begin{center}
		\includegraphics[width=1\columnwidth]{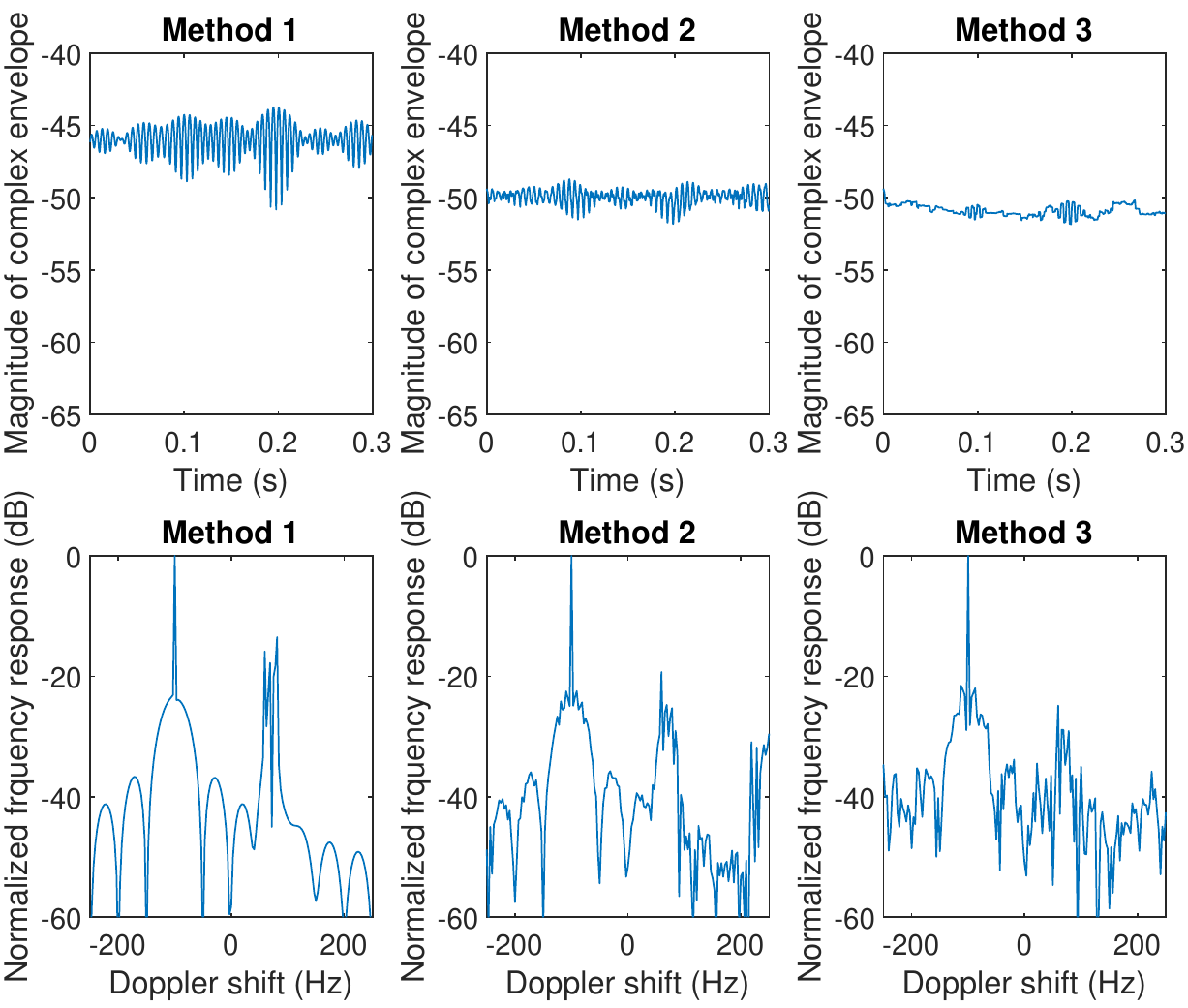}
		\vspace*{-0.4cm}\caption{Complex envelope magnitude and Doppler spectrum for the general case with 10 IOs and $N=M=5$ (Setup I).}\vspace*{-0.3cm}
		\label{fig:Fig19}
	\end{center}
\end{figure}

\begin{figure}[!t]
	\begin{center}
		\includegraphics[width=1\columnwidth]{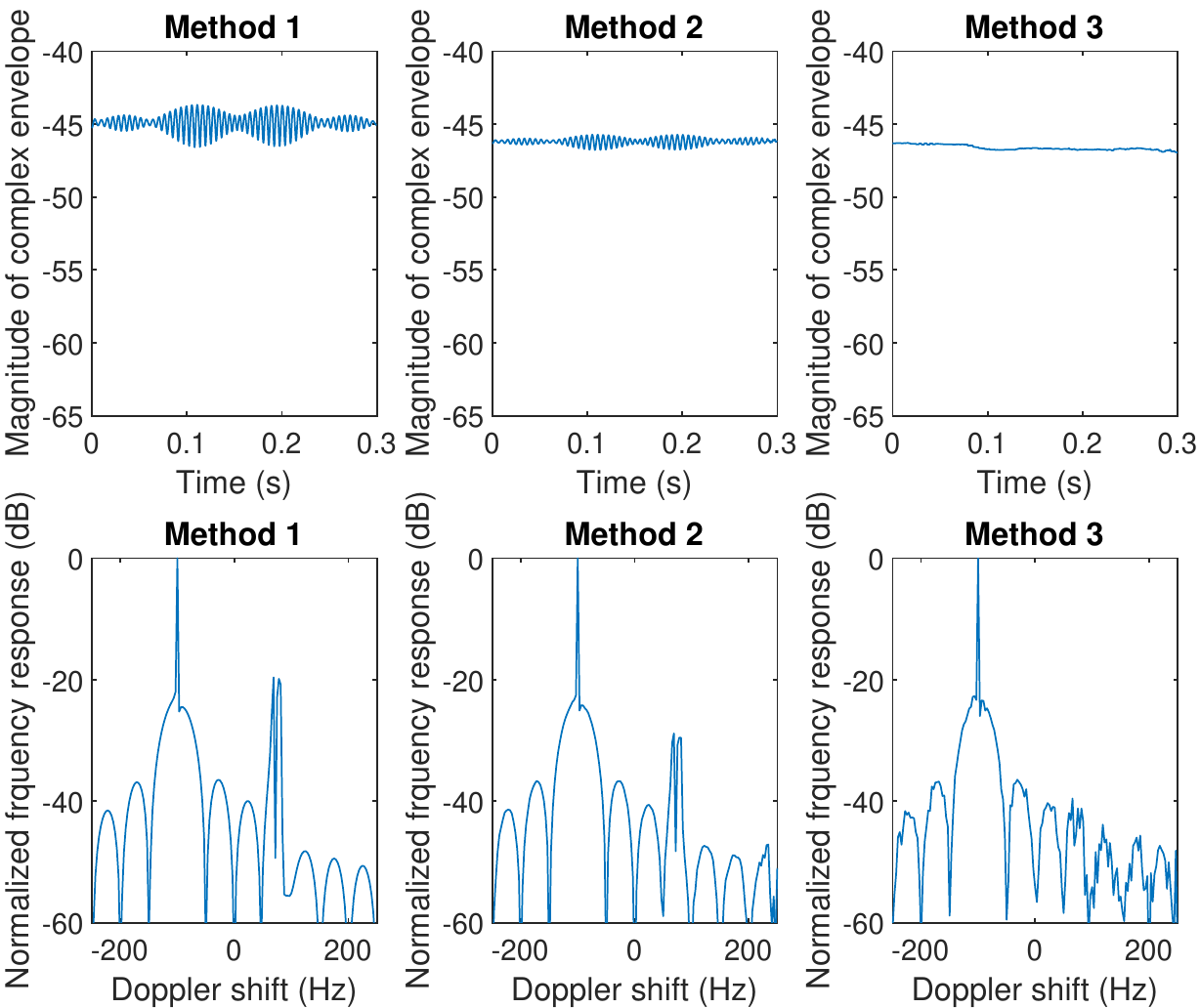}
		\vspace*{-0.4cm}\caption{Complex envelope magnitude and Doppler spectrum for the general case with 10 IOs and $N=7,M=3$ (Setup II).}\vspace*{-0.3cm}
		\label{fig:Fig20}
	\end{center}
\end{figure}

In Figs. \ref{fig:Fig18}-\ref{fig:Fig20}, we consider three different scenarios based on the number of RISs in the system: $N=3,M=7$ (Setup I), $N=M=5$ (Setup I), and $N=7,M=3$ (Setup II) and assess the potential of the introduced Methods 1-3. As seen from Figs. \ref{fig:Fig18}-\ref{fig:Fig20}, although Method 1 ensures a high complex envelope magnitude in average with the price of a larger Doppler spread (faster variation in time), Methods 2 and 3 are more effective in reducing the fade patterns observed in the time domain by modifying the Doppler spectrum through the elimination of plain IO signals. Particularly, the improvements provided by Method 3 are more noticeable both in time and frequency domains. For instance, for the case of $N=7,M=3$, Method 3 almost eliminates all Doppler spectrum components stemming from three plain IOs and ensures an approximately constant magnitude for the complex envelope, as seen from Fig. \ref{fig:Fig20}. 

To gain further insights, in Table \ref{TableI}, we provide a quantitative analysis by comparing the peak-to-peak value $\Delta_r$ of $\left| r(t) \right| $ and its time average $\bar{r}$ (both measured in dB) for all methods, i.e., $\Delta_r= \left| r(t) \right|_{\max} - \left| r(t) \right|_{\min}  $ and $\bar{r}=\frac{1}{n_s} \sum_{n=0}^{n_s-1} \left| r(i t_s) \right|  $, where $n_s$ and $t_s$ respectively stand for the total number of time samples and sampling time, which are selected as $n_s=960$ and $t_s=0.3125$ ms for this specific simulation. As observed from Table \ref{TableI}, increasing $N$ noticeably reduces  $\Delta_r$ for all methods, while this reduction is more remarkable for Methods 2 and 3. We also evince that Methods 2 and 3 cause in a slight degradation in  $\bar{r}$ since they utilize RISs to cancel out reflections from plain IOs. Generalizing our discussion from Subection 4.2.1, we claim that Method 1 can be the preferred choice to maximize the (time-averaged) magnitude  of the complex envelope due to the stronger LOS path, however, the complete mathematical proof of this claim is highly intractable. We also observe that Method 2 provides a nice compromise between Methods 1 and 3 by providing a much lower $\Delta_r$ with a close $\bar{r}$ compared to Method 1, while Method 3 ensures the minimum $\Delta_r$.

\begin{table}[t]
	\caption{Comparison of Methods 1-3 in terms of peak-to-peak variation ($\Delta_r$ in dB) and time-average ($\bar{r}$ in dB) of $\left| r(t) \right| $. }\vspace*{0.2cm}
{\small 	\begin{tabular}{cccc}
		
		& Method 1                                                 & Method 2                                                & Method 3                                                \\ \hline
	$ 	N=3 $, $ M=7 $ & \begin{tabular}[c]{@{}l@{}}$\Delta_r=11.78$\\ $\bar{r}=-47.22$\end{tabular} & \begin{tabular}[c]{@{}l@{}}$\Delta_r=5.43$\\ $\bar{r}=-47.76$\end{tabular} & \begin{tabular}[c]{@{}l@{}}$\Delta_r=3.95$\\ $\bar{r}=-50.17$\end{tabular} \\ \hline
	$ 	N=M=5  $   & \begin{tabular}[c]{@{}l@{}}$\Delta_r=7.08$\\ $\bar{r}=-45.99$\end{tabular}  & \begin{tabular}[c]{@{}l@{}}$\Delta_r=3.09$\\ $\bar{r}=-49.93$\end{tabular} & \begin{tabular}[c]{@{}l@{}}$\Delta_r=2.47$\\ $\bar{r}=-50.88$\end{tabular} \\ \hline
	$ 	N=7 $, $ M=3 $ & \begin{tabular}[c]{@{}l@{}}$\Delta_r=2.91$\\ $\bar{r}=-44.90$\end{tabular}  & \begin{tabular}[c]{@{}l@{}}$\Delta_r=1.05$\\ $\bar{r}=-46.20$\end{tabular} & \begin{tabular}[c]{@{}l@{}}$\Delta_r=0.66$\\ $\bar{r}=-46.62$\end{tabular} \\ \hline
	\end{tabular}}
\label{TableI}
\end{table}

\subsection{The General Case with Multiple IOs and without the Direct Signal}
In this section, we revisit the general case of the previous section (Fig. \ref{fig:CaseIV}), however, without the presence of a LOS path. For this case, the received signal with $N$ RISs and $M$ plain IOs can be expressed as follows:

\begin{equation}\label{eq:20}
r(t)= \frac{\lambda}{4\pi} \left( \sum_{i=1}^{N} \frac{e^{j 2\pi f_{R,i}t-j\psi_i+j\theta_i(t)}}{\tilde{d}_{R,i}}  -\sum_{k=1}^{M} \frac{e^{j 2\pi f_{I,k}t-j\phi_k}}{\tilde{d}_{I,k}}\right). 
\end{equation}
Here, the  three methods introduced in Subsection IV.D can be applied with slight modifications. For Method 1, since there is no LOS path, the available RISs in the system can be aligned to the strongest path, which might be from either an RIS or a plain IO and has the shortest radio path distance. For Methods 2 and 3, when $M\ge N$, we use the same procedures as in the LOS case and assign all $N$ RISs to the plain IOs with different purposes. However, when $N>M$, after applying the same permutation selection procedures, we determine the RIS with the strongest path among the  remaining $N-M$ RISs in lieu of the LOS path and align the rest of the RISs ($N-M-1$ ones) to this strongest RIS for each specific permutation. Our methodology has been summarized below:

\subsubsection{Setup I $(M\ge N)$}
\begin{itemize}
	\item Method 1: We align the existing $N$ RISs to the strongest path. If the strongest path belongs to a RIS, whose index is $a$, we have $\theta_i(t) = - 2\pi f_{R,i}t + \psi_i + 2\pi f_{R,a}t - \psi_a \,\,\,(\mathrm{mod}\,\,\,2\pi) $ for $i=1,\ldots,a-1,a+1,\ldots,N$, while $\theta_a(t)=0$. Otherwise, if the strongest path belongs to a plain IO with index $a$, we have $\theta_i(t) = - 2\pi f_{R,i}t + \psi_i + 2\pi f_{I,a}t - \phi_a + \pi \,\,\,(\mathrm{mod}\,\,\,2\pi) $ for $i=1,2,\ldots,N$. Please note that $a=\arg\min_i  \tilde{d}_{R,i} $ if $\min_i \tilde{d}_{R,i}  < \min_k \tilde{d}_{I,k} $ or $a=\arg\min_k  \tilde{d}_{I,k} $, otherwise.
	\item Method 2: The same as  Method 2 in Subsection IV.D for $M\ge N$ except that $r_n(t_0)$ is obtained from \eqref{eq:20} for the $n$th permutation.
	\item Method 3: The same as  Method 3 in Subsection IV.D for $M\ge N$ except that $r_n(t_0)$ is obtained from \eqref{eq:20}  for the $n$th permutation.
\end{itemize}

\subsubsection{Setup II  $(N>M)$}
\begin{itemize}
	\item Method 1: The same as Method 1 given above.
	\item Method 2: We follow the same steps for Method 2 in Subsection IV.D for $N>M$, however, for $n$th permutation, the strongest RIS is selected among the set $\mathcal{S}_n$ (the set of $N-M$ RISs that are not included in the elimination of IO reflections). Denoting the index of this strongest RIS by $a_n$, where $a_n=\arg \min_{i\in\mathcal{S}_n} \tilde{d}_{R,i}$, we have $\theta_{\mathcal{S}_n^i}(t_0) = - 2\pi f_{R,\mathcal{S}_n^i}t_0 + \psi_{\mathcal{S}_n^i} + 2\pi f_{R,a_n}t_0 - \psi_{a_n}\,\,\,(\mathrm{mod}\,\,\,2\pi) $ for $i=1,2,\ldots,N-M$ with $\mathcal{S}_n^i\neq a_n $ and $\theta_{a_n}(t_0)=0$ for this case. The above procedures are repeated for all permutations and the estimates of the received signal samples are obtained as $r_n(t_0)$ from \eqref{eq:20} for $n=1,2,\ldots,P(N,M)$. After the determination of the optimal permutation $\hat{n}$, we obtain the set of RISs targeting the IOs as $\mathcal{R}_{\hat{n}}$ while the set of remaining RISs are given by $\mathcal{S}_{\hat{n}}$. Finally, RIS angles are determined as in Method 2 in Subsection IV.D for $N>M$ with the exception that the phases of the remaining $N-M$ RISs are aligned as $\hat{\theta}_{\mathcal{S}_{\hat{n}}^i}(t_0) = - 2\pi f_{R,\mathcal{S}_{\hat{n}}^i}t_0 + \psi_{\mathcal{S}_{\hat{n}}^i} + 2\pi f_{R,a_{\hat{n}}}t_0 - \psi_{a_{\hat{n}}}\,\,\,(\mathrm{mod}\,\,\,2\pi) $ for $i=1,2,\ldots,N-M$ with $\mathcal{S}_{\hat{n}}^i\neq a_{\hat{n}} $ and $\hat{\theta}_{a_{\hat{n}}}(t_0)=0$. The above procedures are repeated for all time instants. 
	\item Method 3: This method follows the same procedures as that of Method 2 given above, except the determination of the optimum permutation, which is discussed in Subsection IV.D.
\end{itemize}

\begin{figure}[!t]
	\begin{center}
		\includegraphics[width=1\columnwidth]{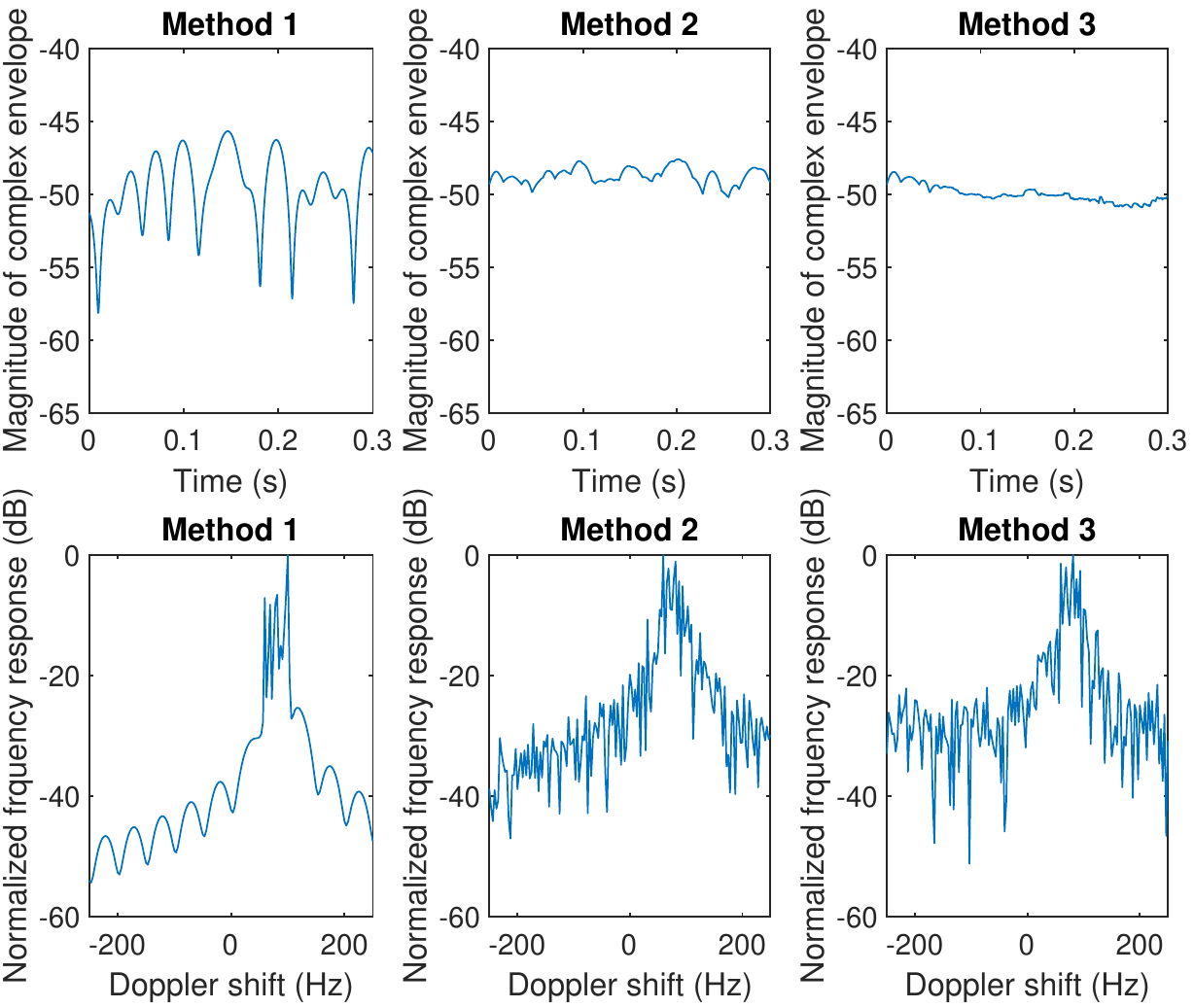}
		\vspace*{-0.4cm}\caption{Complex envelope magnitude and Doppler spectrum for the general case with 10 IOs without a LOS path and $N=3,M=7$ (Setup I).}\vspace*{-0.3cm}
		\label{fig:Fig21}
	\end{center}
\end{figure}

\begin{figure}[!t]
	\begin{center}
		\includegraphics[width=1\columnwidth]{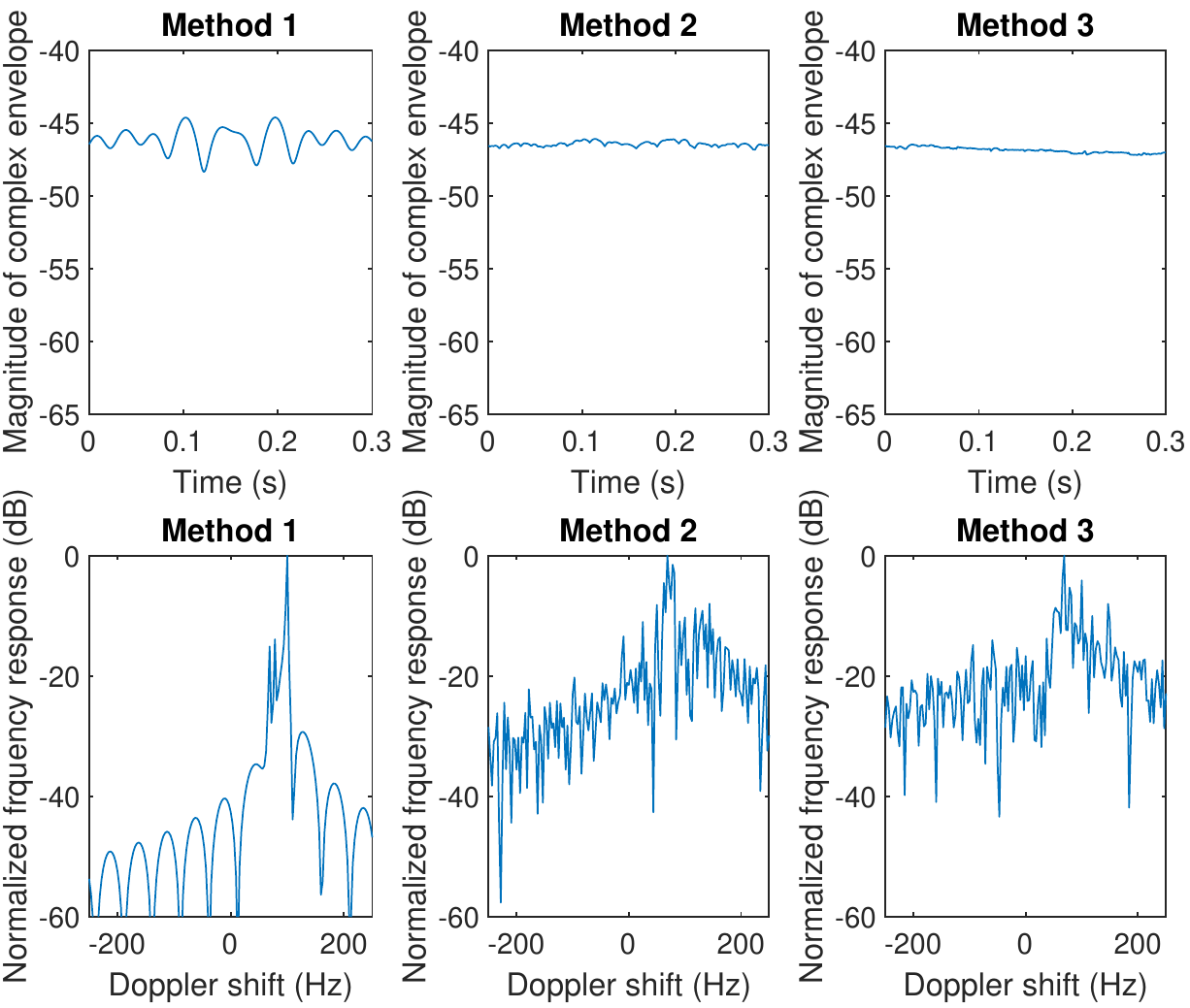}
		\vspace*{-0.4cm}\caption{Complex envelope magnitude and Doppler spectrum for the general case with 10 IOs without a LOS path and $N=7,M=3$ (Setup II).}\vspace*{-0.3cm}
		\label{fig:Fig22}
	\end{center}
\end{figure}

In Figs. \ref{fig:Fig21}-\ref{fig:Fig22}, we investigate the application of Methods 1-3 in two scenarios: $N=3,M=7$ (Setup I) and $N=7,K=3$ (Setup II) for the same simulation scenario of Fig. \ref{fig:Fig16} by ignoring the LOS path. Compared to Figs. \ref{fig:Fig18}-\ref{fig:Fig20}, we observe that due to the nonexistence of the LOS path, all methods provide a similar level of time-average $(\bar{r})$ for the complex envelope while Methods 2 and 3 eliminate deep fades in the received signal. In other words, since we do not have a stronger LOS path, Method 1 loses its main advantage in terms of $\bar{r}$ compared to the other two methods for both scenarios.

It is worth noting that for the case of $N=0$, none of the methods are applicable as in the case of the previous section. However, for $M=0$, Doppler effect can be totally eliminated due to the nonexistence of the LOS path as follows:  $\theta_i(t)=-2 \pi f_{R,i} t+ \phi_i\,\,\,(\mathrm{mod}\,\,\,2\pi)$ for $i=1,2,\ldots,N$.

As a final note, our aim here is to find heuristic solutions to mitigate deep fading and Doppler effects under arbitrary number of RISs and plain IOs, and the determination of the ultimately optimum RIS angles are beyond the scope of this work. Although our methods provide satisfactory results, there might be a certain permutation of RISs/IOs with specific reflection phases that may guarantee a maximized received complex envelope magnitude and/or the lowest Doppler spread. However, the theoretical derivation of this ultimate optimal solution seems intractable at this moment.

\section{Practical Issues}
In this section, we consider a number of practical issues and investigate the performance of our solutions under certain imperfections in the system.

\subsection{Realistic RISs}
\begin{figure}[!t]
	\begin{center}
		\includegraphics[width=1\columnwidth]{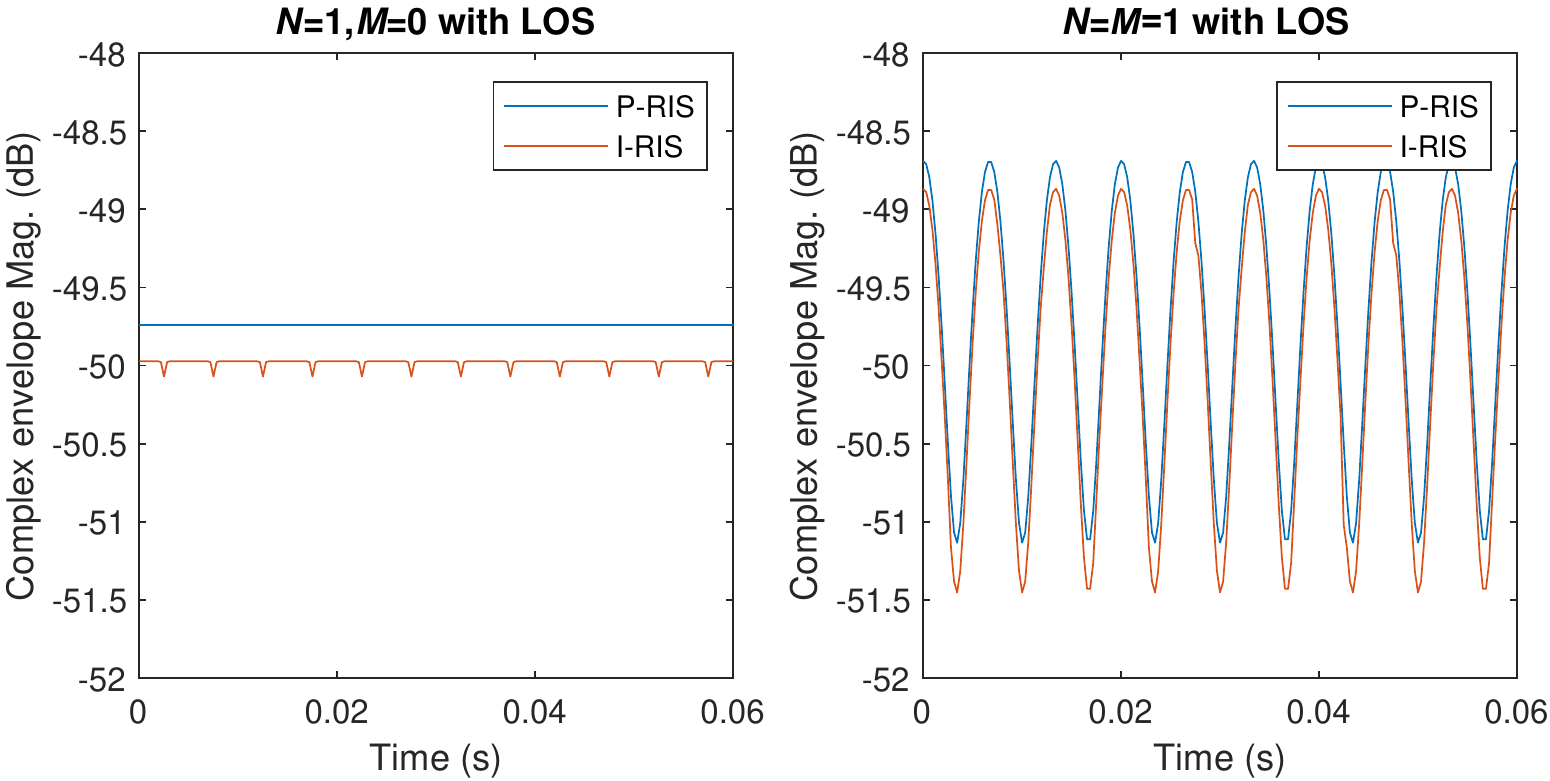}
		\vspace*{-0.45cm}\caption{Complex envelope magnitude in the presence of a realistic RIS for the scenarios of Figs. \ref{fig:CaseI} and \ref{fig:CaseIII}.}\vspace*{-0.3cm}
		\label{fig:Fig23}
	\end{center}
\end{figure}

Throughout this paper, we assumed that the utilized RISs have a unit-amplitude reflection coefficient with a very high resolution reflection phase $\theta(t) \in \left[0,2\pi \right) $ that can be tuned in real time. However, as reported in recent studies, there can be not only a dependency between the amplitude and the phase but also a limited range can be supported for the reflection phase. For this purpose, we consider the realistic RIS design of Tretyakov \textit{et al.} \cite{MDR-16}, which has a reflection amplitude of $-1$ dB with a reflection phase between $-150^{\circ}$ and $140^{\circ}$. In Fig. \ref{fig:Fig23}, we compare the complex envelope magnitudes of two scenarios in the presence of a perfect RIS (P-RIS) and an imperfect RIS (I-RIS) with practical constraints: i) the scenario of Fig. \ref{fig:CaseI} with $N=1,M=0$ and ii) the scenario of Fig. \ref{fig:CaseIII} with $N=M=1$. As seen from Fig. \ref{fig:Fig23}, the practical RIS of \cite{MDR-16} causes a slight degradation both in magnitude and shape of the complex envelope, however, its overall effect is not significant. A further degradation would be expected in the presence of discrete phase shifts \cite{Wu_2018_2}, and this analysis is left for interested readers.

\begin{figure}[!t]
	\begin{center}
		\includegraphics[width=1\columnwidth]{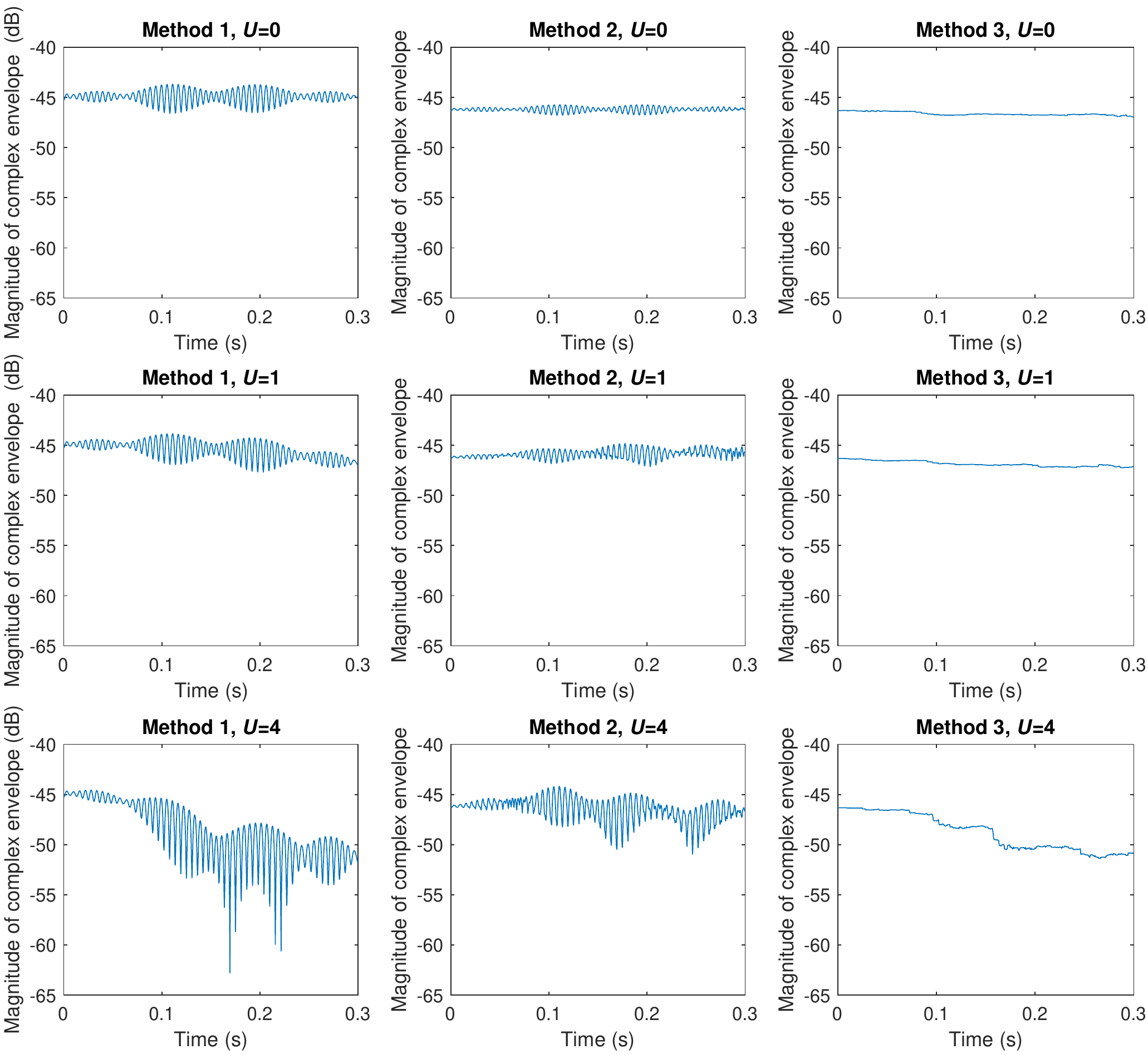}
		\vspace*{-0.4cm}\caption{Complex envelope magnitude for the general case with 10 IOs with a LOS path and $N=7,M=3$ under erroneous Doppler frequency shifts at RISs ($U=1$ and $4$) with the perfect case $(U=0)$.}\vspace*{-0.3cm}
		\label{fig:Fig24}
	\end{center}
\end{figure}  

\subsection{Imperfect Knowledge of Doppler Frequencies}
As discussed in Section IV, in case of multiple RISs, a central processing unit needs to acquire the knowledge of Doppler frequencies of all incoming rays to initiate Methods 1-3 in coordination with the available RISs. Here, we assume that due to erroneous estimation of the velocity of the MS and/or relative positions of the IOs, the RISs in the system are fed back with erroneous Doppler shifts (in Hz), given by $f_{R,i}^e=f_{R,i}+e_{R,i}$ and $f_{I,k}^e=f_{I,k}+e_{I,k}$, while the dominant Doppler shift $(f_D)$ stemming from the LOS path is perfectly known. Here, $e_{R,i}$ and $e_{I,k}$ respectively stand for the errors in Doppler shifts for $i$th RIS and $k$th plain IO. To illustrate the effect of this imperfection, these estimation error terms are modelled by independent and identically distributed uniform random variables in the range $[-U,U]$ (in Hz). In Fig. \ref{fig:Fig24}, we consider the scenario of $N=7,M=3$ with $U=0$, $1$ and $3$ for the same geometry of Fig. \ref{fig:Fig16}. As seen from Fig. \ref{fig:Fig24}, while the degradation in the complex envelope is not a major concern for $U=1$, a significant distortion has been observed for the case of $U=4$ with respect to time. Here, Methods 2 and 3 appear more reliable in the presence of Doppler frequency estimation errors, however, we observe that the overall system is highly sensitive to this type of error.

\subsection{High Mobility \& Discrete-Time RIS Phases}
\begin{figure}[!t]
	\begin{center}
		\includegraphics[width=7.5cm,height=5.5cm]{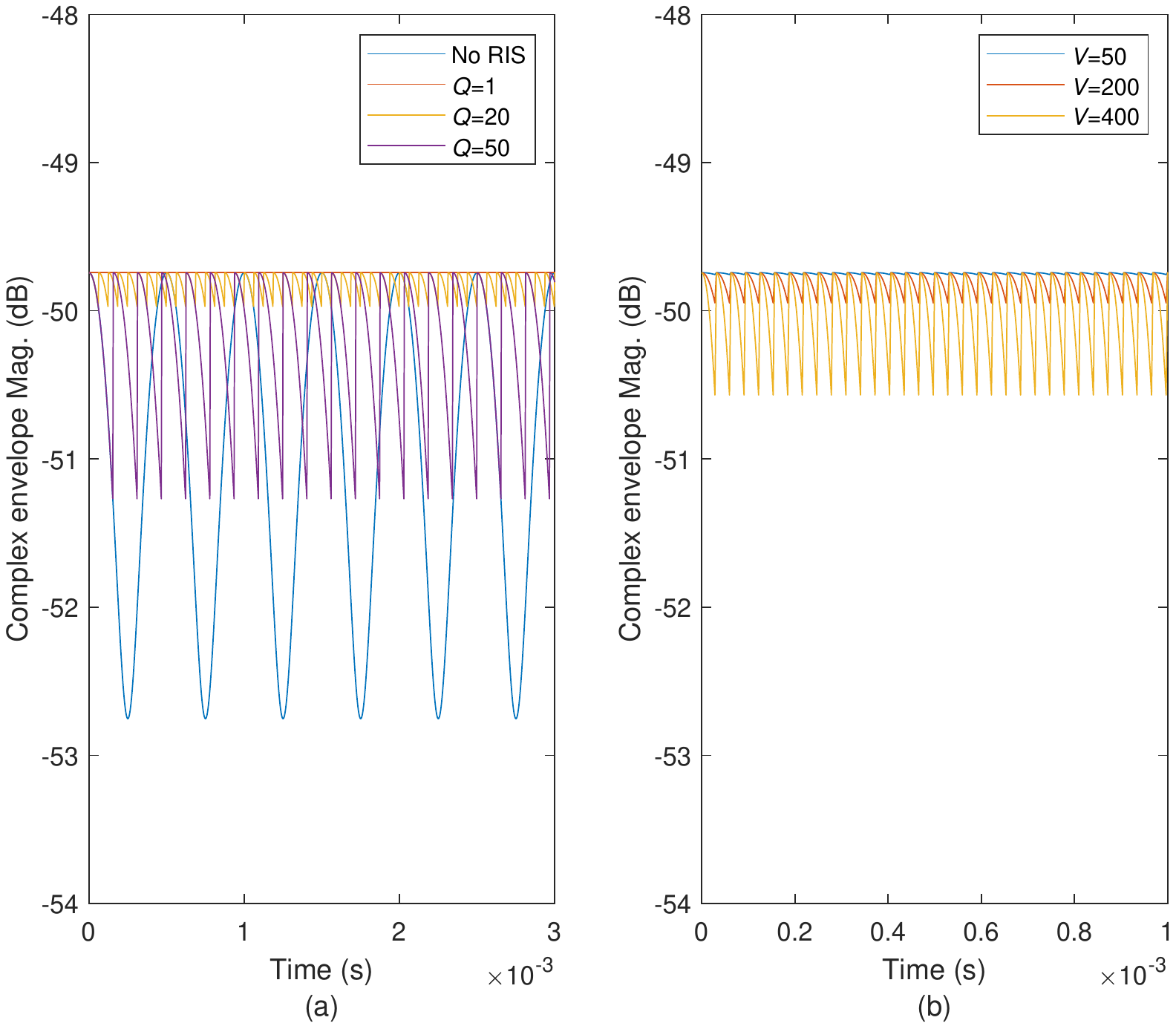}
		\vspace*{-0.4cm}\caption{Complex envelope magnitude for the scenario of Fig. \ref{fig:CaseI} $(N=1,M=0)$ a) under high mobility ($V=100$ m/s) and fixed reflection phases for a period of $Qt_s$ seconds with $Q=1$, $20$, and $50$, b) under increasing Doppler frequencies and a  reflection phase update duration of $t_r=12.5$ $\mu\text{s}$.}\vspace*{-0.3cm}
		\label{fig:Fig25}
	\end{center}
\end{figure}

In this subsection, we will focus on the case of high mobility under the assumption of discrete-time RIS reflection phases. In this scenario, the RIS reflection phases remain constant for a certain time duration. It is worth noting that all methods described earlier are also valid for the case of high mobility if the RIS reflection phases can be tuned in real-time with a sufficiently high rate. However, in practice, due to limitations in terms of the RIS design and signaling overhead in the network, the RIS reflection phases can be tuned at only (certain) discrete-time instants. Let us denote the RIS reconfiguration interval by $t_r$ (in seconds), i.e., the RIS phases can be adjusted in every $t_r$ seconds only. In our first computer simulation, we consider that the complex envelope is represented by its samples taken at every $t_s$ seconds. Here, we assume that once the RIS reflection phases are adjusted according to the LOS path, they remain fixed for $Qt_s$ seconds. In other words, for $Q=1$, we update the RIS reflection phases at each sampling time and obtain the results given throughout the paper. In Fig. \ref{fig:Fig25}(a), we perform this simulation for the high mobility case of $V=100$ m/s, $f_c=3$ GHz and $t_s=3.125$ $\mu\text{s}$ with $N=1$ and $M=0$ (for the basic scenario of Fig. \ref{fig:CaseI}). Here, $t_s$ has been intentionally reduced to capture the variations in the complex envelope with respect to time due to the higher Doppler spread of the unmodulated carrier and a travel distance of $3\lambda$ is considered.  In this case, we assume that RIS reflection phases are modified as $\theta(t)=-4 \pi f_D t\,\,\,(\mathrm{mod}\,\,\,2\pi)$ in every $Qt_s$ seconds, i.e., the RIS cannot be reconfigured fast enough compared to the sampling frequency (variation) of the complex envelope. As seen from Fig. \ref{fig:Fig25}(a), a distortion is observed in the complex envelope due to the delayed reconfiguration of RIS reflection phases. However, we conclude that even if with $Q=50$, the variation in the complex envelope is not as significant as in the case without an RIS (shown in the figure as a benchmark), while the variation is not significant for $Q=20$. In what follows, we present a theoretical framework to describe this phenomenon.

In mathematical terms, for the considered scenario that is formulated by \eqref{eq:4} in terms of its received complex envelope, assuming that the RIS reflection phase is adjusted and fixed at time instant $t_1$ while focusing on the complex envelope at time $t_2>t_1$, we obtain
\begin{align}\label{eq:son}
r(t_2)&= \frac{\lambda}{4\pi}\left(  \frac{e^{ -j 2\pi f_D t_2}}{d_{\text{LOS}}}  + \frac{e^{ j 2\pi f_D t_2+j\theta(t_1)}}{d_{\text{LOS}}+2d_1}    \right) \nonumber \\
&=\frac{\lambda e^{ -j 2\pi f_D t_2} }{4\pi} \left(\frac{1}{d_{\text{LOS}}} +\frac{e^{j 4 \pi f_D \Delta t}}{d_{\text{LOS}}+2d_1} \right)
\end{align}
where $\Delta t=t_2-t_1 <t_r$. Here, we considered the fact that the RIS reflection phase is fixed at time $t_1$ as $\theta(t_1)=-4 \pi f_D t_1$. As a result, we observe a variation in the complex envelope magnitude, which is a function of both $f_D$ and $\Delta t$. It is worth noting that letting $\Delta t=0$ in \eqref{eq:son}, one can obtain \eqref{eq:5} for $t=t_2$. After simple manipulations, the magnitude of the complex envelope is calculated as
\begin{align}\label{eq:son2}
&\left| r(t_2)\right| \nonumber \\
&=   \left( \frac{\lambda}{4\pi}\right) \!\! \left( \frac{1}{d_{\text{LOS}}^2} + \frac{1}{(d_{\text{LOS}}+2d_1)^2} +\frac{2 \cos(4\pi f_D \Delta t)}{d_{\text{LOS}}(d_{\text{LOS}}+2d_1)} \right)^{\!\!1/2}\!\!\!\!.   
\end{align} 
It is evident from \eqref{eq:son2} that the magnitude of the complex envelope is no longer constant unless $4 \pi f_D \Delta t \ll 1 $. In light of the above analysis, to ensure a constant magnitude for the complex envelope, that is, to eliminate the fade pattern due to Doppler spread, we must have $t_r < \frac{1}{40 \pi f_D}$ for the considered scenario. In other words, the RIS should be tuned fast enough compared to $f_D$ to capture the variations of the received signal. To illustrate this effect, in Fig. \ref{fig:Fig25}(b), for a fixed $t_r$ value of $12.5$ $\mu\text{s}$, we change the velocity of the MS and observe the magnitude of the complex envelope. As seen from Fig. \ref{fig:Fig25}(b), while the smaller Doppler frequency of $500$ Hz ($V=50$ m/s) can be captured by the RIS since $t_r < \frac{1}{40 \pi f_D}=15.91$ $\mu\text{s}$  for this scenario, we observe an oscillation in the magnitude for the higher Doppler frequencies of $2$ kHz ($V=200$ m/s) and $4$ kHz ($V=400$ m/s) since the condition of $t_r < \frac{1}{40 \pi f_D} $ is no longer satisfied. In light of the above discussion, we conclude that increasing Doppler frequencies poses a much bigger challenge for the real-time adjustment of RIS reflection phases.

Finally, it is worth noting that in case of slow fading $(1/f_D \gg T_s)$, where $T_s$ is the symbol duration, the channel may be assumed to be static over one or several transmission intervals and the variations in the magnitude of the complex envelope from symbol to symbol (in our case, for unmodulated cosine signals) can be compensated by adjusting RIS reflection phases at every $T_s$ seconds (with slight variations in magnitude if $T_s> \frac{1}{40 \pi f_D}$). On the other hand, in the case of fast fading $(1/f_D <T_s)$, since the channel impulse response changes rapidly within the symbol duration, in order to compensate Doppler and fading effects, i.e., to obtain a fixed magnitude for the complex envelope during a symbol duration, RIS reflection phases should be tuned at a much faster rate compared to $T_s$. As an example, consider the transmission of an unmodulated cosine signal for a period of $3$ ms as in Fig. \ref{fig:Fig25}(a). For this case, we have fast fading due to the large Doppler spread, and this can be eliminated by adjusting the RIS reflection phases at a much faster rate compared to $3$ ms, i.e., $t_r < 7.96$ $\mu\text{s}$. Failure of doing this causes variations in the complex envelope magnitude as shown in Fig. \ref{fig:Fig25}(a).

\section{Conclusions and Future Work}
In this paper, we have revisited the multipath fading phenomenon of mobile communications and provided unique solutions by utilizing the emerging concept of RISs in the presence of Doppler effects. By following a bottom-up approach, first, we have investigated simple propagation scenarios with a single RIS and/or a plain IO. Then we have developed several novel methods for the case of multiple RISs and plain IOs depending on the their total numbers as well as the presence of the LOS path. Finally, we have considered a number of practical issues, including erroneous estimation of Doppler shifts, practical reflection phases, and discrete-time reflection phases, for the target setups and evaluated the overall performance under these imperfections. One of the most important conclusions of this paper is that the multipath fading effect caused by the movement of the mobile receiver/transmitter can be effectively eliminated and/or mitigated by real-time tuneable RISs. A number of interesting trade-offs have been demonstrated between fade pattern elimination and complex envelope magnitude maximization. While this work sheds light on the development of RIS-assisted mobile networks, exploration of amplitude/phase modulations and more practical path loss/propagation models appear as interesting future research directions.

\section*{Appendix}
The received complex envelope in \eqref{eq:12} can be expressed as
\begin{equation}\label{eq:A1}
r(t)= r_{\text{LOS}} e^{j \xi_{\text{LOS}}(t)} + r_{1} e^{j \xi_1(t)} + r_{2} e^{j \xi_2(t)}
\end{equation}
where magnitude and phase values of the LOS and two reflected signals (from IO 1 (RIS) and IO 2) are shown by $r_{\text{LOS}}$, $r_1$, $r_2$ and $\xi_{\text{LOS}}(t)$, $\xi_1(t)$, $\xi_2(t)$, respectively. Here, we are interested in the maximization of $\left|r(t) \right| $ with respect to $\xi_1(t)=2\pi f_D t + \theta_1(t)$, which captures the reconfigurable reflection phase of the RIS. We use the following trigonometric identity: For $z_1=r_1e^{j\xi_1}$, $z_2=r_2e^{j\xi_2}$, $z_3=r_3e^{j\xi_3}$, and $z_4=z_1+z_2+z_3=r_4e^{j\xi_4}$, we have $r_4=(r_1^2 + r_2^2 + r_3^2 +2r_1r_2\cos(\xi_1-\xi_2) +2r_1r_3\cos(\xi_1-\xi_3)+2r_2r_2\cos(\xi_2-\xi_3))^{1/2}$. In light of this, the maximization of $\left|r(t) \right| $ can be formulated as
\begin{align}\label{eq:A2}
&\max_{\theta_1(t)}\,\, \left|r(t) \right|^2  \nonumber \\
&\max_{\theta_1(t)}\,\, r_{\text{LOS}} r_1 \cos(\xi_{\text{LOS}}(t)-\xi_1(t)) + r_1 r_2 \cos(\xi_1(t)-\xi_2(t)) \nonumber \\
&\max_{\theta_1(t)}\,\, r_{\text{LOS}} \cos(4\pi f_D t +\theta_1(t)) \nonumber \\
&\hspace{1cm}+ r_2 \cos(2\pi f_D t(1-\cos\alpha)+\phi_2+\theta_1(t))
\end{align}
where the constant magnitude terms and the term does not contain $\theta_1(t)$ is dropped. Using the identity $\cos(x+y)=\cos x \cos y - \sin x \sin y$ and grouping the terms with $\theta_1(t)$, we obtain
\begin{align}\label{eq:A3}
&\max_{\theta_1(t)}\,\, A\cos\theta_1(t)  + B\sin\theta_1(t) \nonumber \\
&\max_{\theta_1(t)}\,\, \mathrm{sgn}(A)\sqrt{A^2 + B^2} \cos(\theta_1(t) +\tan^{-1}(-B/A)) 
\end{align}
where $A$ and $B$ are as defined in \eqref{eq:15} and the harmonic addition theorem \cite{Harmonic} is used. Consequently, to maximize the complex envelope, we have to ensure 
\begin{equation}\label{eq:A4}
\mathrm{sgn}(A)\cos(\theta_1(t) +\tan^{-1}(-B/A))=1.
\end{equation}
This can be satisfied by
\begin{equation}\label{eq:A5}
\theta_1(t)= \frac{\pi}{2}(1-\mathrm{sgn}(A))- \tan^{-1}(-B/A)
\end{equation}
which completes the proof.

\bibliographystyle{IEEEtran}
\bibliography{bib_2020}

\end{document}